\documentclass[11pt]{article}
\usepackage{natbib,graphicx,dsfont,times}
\usepackage[small]{caption}
\usepackage{amssymb,amsmath}
\usepackage{color}
\usepackage{multirow}
\usepackage{color, colortbl}
\definecolor{Gray}{gray}{0.9}

\begin{document}

\title{A Fully Nonparametric Modeling Approach to Binary Regression}

\author{Maria DeYoreo and Athanasios Kottas
\thanks{M. DeYoreo (mdeyoreo@ams.ucsc.edu) is Ph.D. student, and A. Kottas
(thanos@ams.ucsc.edu) is Professor of Statistics, Department of Applied Mathematics
and Statistics, University of California, Santa Cruz, CA 95064, USA.
}
}

\markboth{DeYoreo and Kottas}{A Fully Nonparametric Modeling Approach to Binary Regression}
\maketitle

\begin{abstract}
\noindent

We propose a general nonparametric Bayesian framework for binary regression, which is built from 
modeling for the joint response-covariate distribution.  The observed binary responses are assumed 
to arise from underlying continuous random variables through discretization, and we model the joint 
distribution of these latent responses and the covariates using a Dirichlet process mixture of 
multivariate normals.  We show that the kernel of the induced mixture model for the observed data 
is identifiable upon a restriction on the latent variables. To allow for appropriate dependence 
structure while facilitating identifiability, we use a square-root-free Cholesky decomposition of 
the covariance matrix in the normal mixture kernel.  In addition to allowing for the necessary 
restriction, this modeling strategy provides substantial simplifications in implementation of 
Markov chain Monte Carlo posterior simulation. 
We present two data examples taken from areas for which the methodology is especially 
well suited. In particular, the first example involves estimation of relationships between 
environmental variables, and the second develops inference for natural selection surfaces 
in evolutionary biology. Finally, we discuss extensions to regression settings with 
multivariate ordinal responses.

\end{abstract}

\noindent


\newpage

\section{Introduction} 
Binary responses measured along with covariates are present in several problems in science and 
engineering. From a modeling perspective, interest centers on determining the regression 
relationship between the response and covariates. 
Standard approaches to this problem -- both classical and Bayesian -- involve potentially 
restrictive distributional assumptions as well as those of linearity in relating the response to the covariates.  
Common modeling techniques result in a small range of monotonic, symmetric trends 
for the probability response curve,
and assume that covariate effects are additive. 

There has been substantial effort devoted to relaxing the functional form 
of the linear predictor through the use of basis functions, including spline 
based approaches \citep[see][for a review of these ideas]{denison}, 
and generalized additive models \citep{hastie}. The latter modify the 
linear predictor by applying a smoothing function to each covariate separately 
and assuming the transformed covariates are additive in their effects. However, 
the underlying distributional assumption is still present through the link function. 

The motivation for Bayesian nonparametric 
methodology lies in the notion that the model should support a wide
range of distributional shapes and regression relationships.
In an effort to create more flexible models to combat overdispersion and asymmetry, 
which the standard links can not, several Bayesian semiparametric binary regression 
methods have been developed.  Early work has targeted either the link, treating it as a 
random function with a nonparametric prior \citep{basu,newton}, or linearity, for instance, 
by viewing the intercept of the linear predictor as arising from an unknown distribution 
\citep{follmann,mukho,walker}. More recently, 
\citet{choudhuri} relaxed the linearity assumption by placing a Gaussian process prior on 
the argument of the inverse link. \citet{trippa} assumed each binary response to arise 
from a random colored tessellation, and placed a Dirichlet process (DP) prior 
\citep{ferguson} on the space of colored tessellations.

\citet{shahbaba}, \citet{dunson}, and \citet{hannah} have 
proposed nonparametric solutions to the regression problem with categorical responses.
These approaches build off the work of \citet{muller}, which modeled the joint distribution of 
continuous responses $y$ and covariates $x$ with a DP mixture of normal 
distributions, inducing a flexible model for $\mathrm{E}(y\mid x)$.  The idea of inducing a regression 
model through the joint response-covariate distribution is attractive, since in many settings 
the covariates 
are not fixed prior to sampling, including several applications in the environmental, 
biomedical, and social sciences.

We target problems of this type, proposing a flexible model for fully nonparametric binary regression, 
in which the responses and covariates arise together as random vectors, requiring a model for their 
joint distribution.
The foundation of the proposed methodology is different from the existing nonparametric modeling approaches. We elaborate further in Section 4, but here note that a key feature of the proposed model involves the introduction of latent continuous responses, in similar spirit to parametric probit models; see, for instance, \citet{albert}.  Let $\{(y_{i},x_{i}):i=1,...,n\}$ denote the data, where each observation consists of a binary response $y_{i}$ along with a vector of covariates,  $x_{i}=(x_{i1},...,x_{ip})$. The continuous auxiliary variables, $z_i$, determine the observed binary responses $y_i$ by their sign, such that $y_i=1$ if and only if $z_i>0$. Instead of seeking a nonparametric model for the regression function, we estimate the joint distribution of latent responses and covariates, $f(z,x)$, using a 
DP mixture of multivariate normal distributions, which induces a flexible 
model for the regression relationship, $\mathrm{Pr}(y=1\mid x)$. In addition to providing a general modeling platform, the latent responses are 
conceptually meaningful in many applications. 
The proposed model is shown to be identifiable provided the variance of $z$ 
within each mixture component is fixed, a restriction implemented through a 
square-root-free Cholesky decomposition of the mixture kernel covariance matrix. 
This aspect of the model formulation retains computational efficiency in posterior
simulation while enabling the use of priors more flexible than the inverse-Wishart distribution. 
We develop two approaches to prior specification for the covariance parameters, one which 
involves prior simulation and can be used for problems with a small number of covariates, 
and a second which is 
more straightforward to apply as the dimension of the covariate space increases.

In Section 2, we formulate the mixture model for binary regression.  
We discuss identifiability for the parameters of the mixture kernel distribution, 
as well as prior specification approaches, and give details for posterior inference.  
In Section 3, the methodology is applied to problems from environmetrics and evolutionary 
biology, using two data sets from the literature for illustration. 
The latter example involves estimation of fitness surfaces, a problem for which our 
method is particularly powerful relative to existing approaches.
Section 4 contains further discussion to place our contribution within the existing 
literature, and to indicate possible extensions. Technical details on the identifiability 
result, prior specification and posterior simulation, and the expressions for the model 
comparison criterion used in Section 3 are provided in the appendices.

\section{Methodology}

\subsection{The modeling approach}

Focusing on $p$ continuous covariates, $x=(x_{1},...,x_{p})$, and a single binary response $y$, with corresponding latent continuous response $z$, a normal distribution is a natural choice for the kernel in a mixture representation for $f(z,x)$.  
The DP is then used as a prior for the random mixing distribution $G$, to create a mixture model of the form: $f(z,x;G)=\int \mathrm{N}_{p+1}(z,x;\mu,\Sigma)\mathrm{d}G(\mu,\Sigma)$, $G\mid \alpha,\psi \sim \mathrm{DP}(\alpha,G_{0}(\cdot;\psi))$, 
where $\alpha$ is the DP precision parameter, and $\psi$ the parameters 
of the DP centering distribution.

According to the DP constructive definition \citep{sethuraman}, a 
$\mathrm{DP}(\alpha,G_{0})$ realization $G$ is almost surely of the form 
$\sum_{l=1}^{\infty}p_{l}\delta_{\nu_{l}}$, with $\nu_{l}$ independent realizations 
from $G_{0}$, and $p_{l}$ arising through stick-breaking from beta random variables.  
In particular, let $\zeta_{m}$ be independent 
$\mathrm{beta}(1,\alpha)$, $m=1,2,...,$ and define $p_{1}=\zeta_{1}$, and $p_{l}=$
$\zeta_{l}\prod_{r=1}^{l-1}(1-\zeta_{r})$, for $l=2,3,...$; moreover, $\{ \zeta_{m}:m=1,2,\dots\}$ 
and $\{\nu_{l}:l=1,2,\dots\}$ are independent sequences of random variables.  
Applying the constructive definition with $\nu_{l}=(\mu_{l},\Sigma_{l})$, the model admits 
a representation as a countable mixture of multivariate normals, 
$f(z,x;G)=$ $\sum_{l=1}^{\infty}p_{l}\mathrm{N}_{p+1}(z,x;\mu_{l},\Sigma_{l})$.

For the normal kernel distribution, let $\mu^{z}$ denote the mean of $z$, ${\mu^{x}}$ denote the mean of $x$, and partition the covariance matrix such that $\Sigma^{zz}=\mathrm{var}(z)$, ${\Sigma^{xx}}=\mathrm{cov}(x)$, a $p\times p$ matrix, and $\Sigma^{zx}=\mathrm{cov}(z,x)$, a row vector of length $p$.  Then, integrating over the latent response $z$, the induced model for the observables assumes the form
\begin{equation}\label{eqn:inducedmodel}
f(y,x;G)=\sum_{l=1}^{\infty}p_{l}\mathrm{N}_{p}(x;{\mu_{l}^{x}},{\Sigma_{l}^{xx}})\mathrm{Bern}\left(y;\Phi \left(\frac{\mu_{l}^{z}+{\Sigma_{l}^{zx}}({\Sigma_{l}^{xx}})^{-1}
(x-{\mu_{l}^{x}})}{(\Sigma_{l}^{zz}-{\Sigma_{l}^{zx}}({\Sigma_{l}^{xx}})^{-1}({\Sigma_{l}^{zx}})^{t})^{1/2}}\right)\right),
\end{equation}
where $\Phi(\cdot)$ denotes the standard normal cumulative distribution function.

Flexible inference for the binary regression functional can be obtained through  
$\mathrm{Pr}(y=1\mid x;G)=\mathrm{Pr}(y=1,x;G)/f(x;G)$.  Marginalizing over $z$ in 
$f(z,x;G)$, the marginal distribution for $x$ is $f(x;G)=$
$\sum_{l=1}^{\infty}p_{l}\mathrm{N}_{p}(x;{\mu_{l}^{x}},{\Sigma_{l}^{xx}})$. 
Hence, the implied conditional regression function can be expressed as a 
weighted sum of the form $\sum_{l=1}^{\infty}w_{l}(x)\pi_{l}(x)$, with covariate-dependent weights $w_{l}(x)=p_{l}\mathrm{N}_{p}(x;{\mu_{l}^{x}},{\Sigma_{l}^{xx}})/\sum_{j=1}^{\infty}p_{j}\mathrm{N}_{p}(x;{\mu_{j}^{x}},{\Sigma_{j}^{xx}})$, and probabilities 
\begin{equation}\label{eqn:piweights}\pi_{l}(x)=\Phi \left(\frac{\mu_{l}^{z}+{\Sigma_{l}^{zx}}({\Sigma_{l}^{xx}})^{-1}(x-{\mu_{l}^{x}})}{(\Sigma_{l}^{zz}-{\Sigma_{l}^{zx}}({\Sigma_{l}^{xx}})^{-1}({\Sigma_{l}^{zx}})^{t})^{1/2}}\right),\end{equation} 
which have the probit form with component-specific intercept and slope parameters.

The dependence structure of the mixture kernel in $f(z,x;G)$
is key to obtaining general inference for the implied binary regression function. 
However, is it sensible to leave all elements of the kernel covariance matrix $\Sigma$ unrestricted? 
In the case of a single mixture component, which arises in the limit as $\alpha\rightarrow0^{+}$, 
the regression function $\mathrm{Pr}(y=1\mid x;G)$ has the form a single normal cumulative 
distribution function, as given in (\ref{eqn:piweights}). This function takes the same value for 
any $x$ when $\mu^z$ and $\Sigma^{zx}$ are scaled by a positive constant $c$, and $\Sigma^{zz}$ 
by $c^2$, indicating that different combinations of $\mu$ and $\Sigma$ result in the same 
probability of positive response.
Hence, there is an identification problem if $\mu$ and $\Sigma$ are 
unrestricted. This limiting case of our model is a parametric probit model, 
albeit with random covariates. In this setting, if identification constraints are not imposed,
then prior distributions become increasingly important yet difficult to specify, and the use of 
noninformative priors can be problematic and create computational difficulties 
\citep{hobert, mcculloch, koop}. In addition, 
empirical evidence based on simulated data suggests 
that, without parameter restrictions, the correlations implied by the covariance matrices 
$\Sigma_l$ are not representative of the correlations that generated the data, and 
undesirable behavior is present in the uncertainty bands of the binary regression functional 
at the extreme regions of the covariate space. For these reasons, and the fundamental belief 
that within a particular cluster or mixture component the corresponding parameters should 
be identifiable, we now focus on restricting the kernel of the mixture.

Here, we employ the standard definition of likelihood identifiability, such that a parameter $\theta$ for a family of distributions $\{f(x\mid \theta):\theta \in \Theta\}$ is identifiable if distinct values of $\theta$ correspond to distinct probability density functions, that is, if $\theta\neq \theta'$, then $f(x\mid\theta)$ is not the same function of $x$ as $f(x\mid\theta')$.  Under our setting, the focus is on the kernel of the mixture model for the observed data, $f(y,x;G)$, which has the form 
\begin{equation}\label{eqn:induced}\normalsize {k(y,x;\eta)=
\mathrm{N}_{p}(x;{\mu^{x}},{\Sigma^{xx}})\mathrm{Bern}\left(y;\Phi \left(\frac{\mu^{z}+{\Sigma^{zx}}({\Sigma^{xx}})^{-1}(x-{\mu^{x}})}
{(\Sigma^{zz}-{\Sigma^{zx}}({\Sigma^{xx}})^{-1}({\Sigma^{zx}})^{t})^{1/2}}\right)\right)},
\end{equation}
with $\eta=({\mu^{x}},\mu^{z},\Sigma^{xx},{\Sigma^{zz}},{\Sigma^{zx}})$. Note that if $z$ and $x$ are independent in the mixture kernel, the probability in the Bernoulli response becomes $\Phi(\mu^{z}/(\Sigma^{zz})^{1/2})$; 
hence, a restriction -- for instance, on $\Sigma^{zz}$ -- is required for identifiability.  
This is in fact the only restriction necessary to obtain an identifiable kernel, and we 
thus retain the ability to estimate ${\Sigma^{zx}}$, which is significant in capturing 
the dependence of $y$ on $x$ under the mixture distribution.  The specific result 
is given in the following lemma whose proof can be found in Appendix A.
\\
\\
LEMMA 1. \emph{The parameters $({\mu^{x}},\mu^{z},{\Sigma^{xx}},{\Sigma^{zx}})$ are identifiable in the model for observed data which has the form in (\ref{eqn:induced}), provided $\Sigma^{zz}$ is fixed to a constant.}
\\

While intuitively straightforward, fixing $\Sigma^{zz}$ to a constant is challenging operationally. The usual conditionally conjugate inverse-Wishart choice for $G_{0}(\Sigma)$ does not offer the solution, due to the single degree of freedom parameter in the inverse-Wishart distribution, which does not allow for one element of $\Sigma$ to be fixed while freely estimating the rest of the matrix.  One way to overcome this problem is to reparameterize $\Sigma$, using a square-root-free Cholesky decomposition.  This decomposition is useful for modeling longitudinal data \citep{daniels}, as well as specifying conditional independence assumptions for the elements of a normal random vector \citep{webb}. Let ${\beta}$ be a unit lower triangular matrix, and let ${\Delta}$ be a diagonal matrix with positive elements, $(\delta_{1},...,\delta_{p+1})$, such that ${\Delta}={\beta\Sigma\beta}^{t}$.  Hence, $\Sigma=\beta^{-1}\Delta(\beta^{-1})^{t}$, where $\beta^{-1}$ is also lower triangular with all its diagonal elements equal to 1, and $\mathrm{det}(\Sigma)=\prod_{i=1}^{p+1}\delta_{i}$. Moreover, $\delta_{1}=\Sigma^{zz}$, 
and thus the identifiability restriction can be implemented by setting the first element of ${\Delta}$ 
equal to a constant value; $\delta_{1} = 1$ is used from this point forward.  Instead of mixing directly 
on $\Sigma$, the mixing takes place on ${\beta}$ and the $p$ free elements of ${\Delta}$, 
$(\delta_{2},...,\delta_{p+1})$.  Hence, the mixture model 
for the joint density of the latent response and covariates is now written as:
\begin{equation}\label{eqn:dpmixbeta}
f(z,x;G) = \sum_{l=1}^{\infty}p_l \mathrm{N}_{p+1}(z,x;{\mu_l},{\beta_l}^{-1}\Delta_l(\beta_l^{-1})^{t}).
\end{equation}

While this decomposition of $\Sigma$ allows for the necessary flexibility in viewing only 
part of the covariance matrix as random, its real utility lies in the existence of a conditionally 
conjugate centering distribution $G_{0}$, which enables development of an efficient Gibbs 
sampler for posterior simulation.  In particular, a multivariate normal $G_0$ component 
for the vector, $\tilde{\beta}$, of $p(p+1)/2=q$ free elements of ${\beta}$, and independent 
inverse-gamma components for $\delta_{2},...,\delta_{p+1}$ result in full conditional 
distributions which are multivariate normal and inverse-gamma, respectively.  
Therefore, $G_{0}$ comprises independent components for ${\mu},\tilde{{\beta}}$, 
and $\delta_{2},...,\delta_{p+1}$, such that it has the form $\mathrm{N}_{p+1}({\mu};{m},{V})\mathrm{N}_{q}(\tilde{{\beta}};{\theta},{C})\prod_{i=2}^{p+1}\mathrm{IG}(\delta_{i};\nu_i,s_{i})$.

\subsection{Posterior inference for binary regression}

In order to simulate from the full posterior distribution, we utilize the blocked Gibbs 
sampler \citep{ishwarenzar, ishjames}.  As a consequence 
of the constructive definition of the DP, any distribution it generates 
can be represented as a countable mixture of point masses.  This definition motivates 
the blocked Gibbs sampler, as it is based on a finite truncation approximation to 
$G$. Specifically, $G$ is truncated to $G_{N}=$ $\sum_{l=1}^{N}p_{l}\delta_{{W}_{l}}$, 
where ${W_{l}}=$ $({\mu}_{l},\tilde{\beta}_{l},{\Delta}_{l})$, and $p_1,\dots,p_{N-1}$ are defined 
through stick-breaking as in the original DP definition, whereas 
$p_N=1-\sum_{l=1}^{N-1}p_l$. Introducing configuration variables $L=$ $(L_{1},...,L_{n})$, 
each taking values in $\{1,...,N\}$, the hierarchical version of the DP 
mixture model for the data given the latent continuous responses, $z=$ $(z_{1},...,z_{n})$, 
becomes
\begin{eqnarray*}
y_{i}\mid z_{i} \stackrel{ind.}{\sim} \mathrm{1}_{(y_{i}=1)}\mathrm{1}_{(z_{i}>0)} + 
\mathrm{1}_{(y_{i}=0)}\mathrm{1}_{(z_{i}\leq 0)}, \quad i=1,...,n 
\nonumber \\
(z_{i},x_{i})\mid{W},L_{i} \stackrel{ind.}{\sim} 
\mathrm{N}_{p+1}(z_{i},x_{i};{\mu}_{L_{i}},{\beta}_{L_{i}}^{-1}{\Delta}_{L_{i}}({\beta}_{L_{i}}^{-1})^{t}), \quad i=1,...,n
\nonumber \\
L_{i} \mid p \stackrel{ind.}{\sim} \sum_{l=1}^{N}p_{l}\delta_{l}(L_{i}), \quad i=1,...,n
\nonumber \\
{W}_{l}\mid\psi \stackrel{ind.}{\sim} \mathrm{N}_{p+1}({\mu}_{l};{m},{V})\mathrm{N}_{q}(\tilde{{\beta}}_{l};{\theta},C)
\prod_{i=2}^{p+1}\mathrm{IG}(\delta_{i,l};\nu_i,s_{i}) \quad l=1,...,N
\nonumber \\
\end{eqnarray*}
where $W=(W_{1},\dots,W_{N})$, and the prior implied for $p=(p_{1},...,p_{N})$ by the stick-breaking construction defined through beta$(1,\alpha)$ random variables corresponds to a generalized Dirichlet distribution (Connor \& Mosimann, 1969). The full Bayesian model is completed with a $\mathrm{gamma}(a_{\alpha},b_{\alpha})$ prior 
for $\alpha$, with mean $a_{\alpha}/b_{\alpha}$, and with conditionally conjugate hyperpriors for ${\psi}=$
$(m,V,\theta,C,s_2,\dots,s_{p+1})$, specifically: $m\sim \mathrm{N}_{p+1}(a_m,B_m)$, 
$V \sim \mathrm{IW}_{p+1}(a_V,B_V)$, $\theta \sim \mathrm{N}_{q}(a_\theta,B_\theta)$, 
$C \sim \mathrm{IW}_{q}(a_C,B_C)$, and 
$s_{i} \stackrel{ind.}{\sim} \mathrm{gamma}(a_{s_{i}},b_{s_{i}})$, for $i=2,...,p+1$.
Here, $S \sim \mathrm{IW}_{k}(a,B)$ indicates that the $k \times k$ positive definite 
matrix $S$ follows an inverse-Wishart distribution with density proportional to
$|S|^{-(a + k + 1)/2} \exp\{ -0.5 \mathrm{tr}(B S^{-1}) \}$.
The notation $\delta_{i,l}$ is used for element $i$ of the vector ${\delta}_{l}$ 
corresponding to the diagonal of ${\Delta}_{l}$. Moreover, where convenient, we use the 
$\Sigma$ notation for the structured covariance matrix, where the elements of $\Sigma$ 
are computed through $\Sigma={\beta}^{-1}{\Delta}({\beta}^{-1})^{t}$.

A key feature of the modeling approach is that simulation from the full posterior 
distribution, $p({W},{L},{p},{\psi},\alpha,{z}|\mathrm{data})$, is possible via Gibbs sampling.
We next discuss posterior simulation details focusing on a result that enables 
Gibbs sampling updates for the parameters that define the covariance matrices 
of the normal mixture components. 

The updates for $p$ and $\alpha$ are generic for any choice of mixture kernel; 
see \citet{ishwarenzar}. Each $L_{i}$, $i=1,...,n$, is sampled from a 
discrete distribution on $\{1,...,N\}$, with probabilities proportional to 
$p_{l}\mathrm{N}_{p+1}(z_{i},x_{i};\mu_{l},\Sigma_{l})$, for $l=1,...,N$. The full 
conditional distributions for the components of ${\psi}$ are easily found 
using standard conjugate updating. 
%
%
The full conditional distribution for each $z_{i}$ is a truncated version of 
the normal distribution $\mathrm{N}(\mu^{z}_{L_{i}} + 
{\Sigma}^{zx}_{L_{i}}({\Sigma}^{xx}_{L_{i}})^{-1}(x_{i}-{\mu}^{x}_{L_{i}}),
1-{\Sigma}^{zx}_{L_{i}}({\Sigma}^{xx}_{L_{i}})^{-1}({\Sigma}^{zx}_{L_{i}})^{t})$, with the 
restriction  $z_{i}>0$ if $y_{i}=1$, and $z_{i}\leq 0$ if $y_{i}=0$.

Letting $\{L_{j}^{*},j=1,...,n^{*}\}$ be the vector of distinct values of ${L}$, the full 
conditional distribution for ${W}_{l}$ is proportional to $G_{0}({W}_{l}\mid{\psi})\prod_{j=1}^{n^{*}}
\prod_{\{i:L_{i}=L_{j}^{*}\}}\mathrm{N}_{p+1}(z_{i},x_{i};\mu_{L_{j}^{*}},{\beta}_{L_{j}^{*}}^{-1}
{\Delta}_{L_{j}^{*}}({\beta}_{L_{j}^{*}}^{-1})^{t})$. If $l\notin\{L_{j}^{*}:j=1,...,n^{*}\}$, 
then ${W}_{l}\sim G_{0}(\cdot\mid\psi)$. If $l\in\{L_{j}^{*}:j=1,...,n^{*}\}$, 
then the full conditional distribution for each element of 
${W}_{l}=$ $({\mu}_{l},\tilde{\beta}_{l},\delta_{2,l},...,\delta_{p+1,l})$ arises from 
the product of a normal likelihood component, based on $\{ (z_{i},x_{i}): L_{i}=L_{j}^{*} \}$,
and the base distribution $G_{0}$.  Therefore, when $l = L_{j}^{*}$, for $j=1,...,n^{*}$, 
the full conditional for $\mu_{l}$ is multivariate normal with mean vector $(V^{-1}+M_{l}\Sigma_{l}^{-1})^{-1}(V^{-1}m+\Sigma_{l}^{-1}\sum_{\{i:L_{i}=l\}}(z_{i},x_{i})^{t})$ 
and covariance matrix $(V^{-1}+M_{l}\Sigma_{l}^{-1})^{-1}$, where $M_{l}=|\{i:L_{i}=l\}|$ 
is the size of mixture component $l$.

Lemma 2, whose proof can be found in Appendix A, provides the result for the 
posterior full conditional distributions of the $\tilde{\beta}_{l}$ and the 
$\delta_{i,l}$, for $i=2,...,p+1$. Before stating the lemma, we fix the required notation.
As discussed earlier, vector $\tilde{{\beta}}$ consists of the lower triangle of free 
elements of matrix ${\beta}$. For instance, if $p=2$, the mixture kernel is a trivariate 
normal, and the free elements of $\beta$ are $({\beta}_{21},{\beta}_{31},{\beta}_{32})$, 
corresponding to $\tilde{{\beta}}=$ $(\tilde{\beta}_{1},\tilde{\beta}_{2},\tilde{\beta}_{3})$. 
The matrix $\Delta$ contains vector $\delta$ on its diagonal. Let $r=$ $p+1$ 
represent the dimension of the mixture kernel. Let ${d}_{i}$ be a vector of length 
$r(r-1)/2=q$, containing $r-1$ nonzero terms, occurring in elements $k(k+1)/2$ for $k=1,...,r-1$. Let ${T}_{i}$ be a block diagonal matrix of dimension $q\times q$ with $r-1$ blocks, which can be constructed from square matrices ${T}_{i}^{1},...,{T}_{i}^{r-1}$ of dimensions $1,...,r-1$. Matrix ${T}_{i}^{j}$ occurs in rows and columns $j(j-1)/2+1$ to $j(j+1)/2$ of ${T}_{i}$.
\\
\\
LEMMA 2. \emph{Consider the following Bayesian probability model:
\[
(y_{i,1},...,y_{i,r})  \mid \mu,\tilde{{\beta}},\delta \stackrel{ind.}{\sim} \mathrm{N}_{r}(\mu,\beta^{-1}\Delta(\beta^{-1})^{t}), 
\,\,\,\,\, i=1,...,n, 
\]
with a multivariate normal prior for $\mu$, independent inverse-gamma priors on the diagonal 
elements of $\Delta$, $\delta_{k}\sim \mathrm{IG}(\nu_{k},s_{k})$, $k=1,...,r$, and a multivariate 
normal prior on the vector comprising the lower triangular elements of $\beta$,
$\tilde{\beta}\sim \mathrm{N}_{q}(\theta,D)$. 
Then, the posterior full conditional distribution for $\delta_{k}$, $k=1,...,r$, is an 
inverse-gamma distribution with shape parameter $\nu_{k}+ 0.5 n$ and scale parameter 
$s_{k} + 0.5 \sum_{i=1}^{n} \{ (y_{i,k}-\mu_{k}) + \sum_{j<k}\beta_{kj}(y_{i,j}-\mu_{j}) \}^{2}$.  
In addition, the posterior full conditional for $\tilde{\beta}$ is multivariate normal with 
mean vector $(D^{-1}+\sum_{i=1}^{n}T_{i})^{-1}(D^{-1}\theta+\sum_{i=1}^{n}T_{i}d_{i})$ and 
covariance matrix $(D^{-1}+\sum_{i=1}^{n}T_{i})^{-1}$. Here, the non-zero elements of $d_{i}$ 
are $-(y_{i,2}-{\mu}_{2})/(y_{i,1}-{\mu}_{1}),...,-(y_{i,r}-{\mu}_{r})/(y_{i,r-1}-{\mu}_{r-1})$, and  
the $(m,n)$-th element of matrix $T_{i}^{j}$, for $j=1,...,r-1$, is given by 
${T}^{j}_{i,mn}=$ $(y_{i,m}-{\mu}_{m})(y_{i,n}-{\mu}_{n})/\delta_{j+1}$, for $m=1,...,j,n=1,...,j$.}
\\
\\
This lemma provides the information necessary to obtain the remaining full conditional 
distributions, which are available in closed form.  Let ${y}_{i}^{*}=$ $(z_{i},x_{i})$ denote 
the augmented latent response-covariate vector, such that ${y}_{i,1}^{*}=$ $z_{i}$
and ${y}_{i,j+1}^{*}=$ $x_{ij}$, for $j=1,...,p$. Then, when $l = L_{j}^{*}$, for $j=1,...,n^{*}$, 
the full conditional distribution for 
${\delta}_{k,l}$ is inverse-gamma with shape parameter $\nu_{k} + 0.5 M_{l}$ and scale parameter 
$s_{k} + 0.5 \sum_{\{i:L_{i}=L_{j}^{*}\}} \{ ({y}^{*}_{i,k}-{\mu}_{k,l}) + 
\sum_{j<k}{\beta}_{kj,l}({y}^{*}_{i,j}-{\mu}_{j,l}) \}^{2}$.
The full conditional for $\tilde{{\beta}}_{l}$ is multivariate normal with covariance 
matrix $(C^{-1} + \sum_{\{i:L_{i}=L_{j}^{*}\}}{T}_{i})^{-1}$, and mean vector $(C^{-1}+\sum_{\{i:L_{i}=L_{j}^{*}\}}{T}_{i})^{-1}(C^{-1}{\theta}+\sum_{\{i:L_{i}=L_{j}^{*}\}}{T}_{i}{d}_{i})$. 
The $p$ non-zero terms in the vector ${d}_{i}$ are $-(y^{*}_{i,2}-{\mu}_{2,l})/(y^{*}_{i,1}-{\mu}_{1,l}),...,-(y^{*}_{i,p+1}-{\mu}_{p+1,l})/(y^{*}_{i,p}-{\mu}_{p,l})$,
and for $j=1,...,p$, the matrix $T_{i}^{j}$ contains elements ${T}^{j}_{i,mn}=$
$(y^{*}_{i,m}-{\mu}_{m,l})(y^{*}_{i,n}-{\mu}_{n,l})/\delta_{j+1,l}$, $m=1,...,j,n=1,...,j$.

The mixing distribution $G$, approximated by $G_N=({p},{W})$, is imputed as a component of the posterior 
simulation algorithm, enabling full inference for any functional of $f(y,x;G)$. 
%
%
The binary regression functional $\mathrm{Pr}(y=1\mid x;G)$ is the main quantity of interest, and is estimated by $\mathrm{Pr}(y=1,x;G)/f(x;G)$, where
%
%
$\mathrm{Pr}(y=1,x;G)=$ 
$\sum_{l=1}^{N}p_{l}\mathrm{N}_{p}(x;{\mu_{l}^{x}},{\Sigma_{l}^{xx}})\pi_{l}(x)$, with $\pi_{l}(x)$ given in (\ref{eqn:piweights}), and 
$f(x;G)=$ $\sum_{l=1}^{N}p_{l}\mathrm{N}_{p}(x;{\mu_{l}^{x}},{\Sigma_{l}^{xx}})$.
Therefore, full inference for $\mathrm{Pr}(y=1\mid x;G)$ can be readily obtained for 
any covariate value $x$, providing a point estimate along with uncertainty 
quantification for the binary regression function.  
Inference can also be obtained for the covariate distribution, $f(x;G)$, as well 
as the covariate distribution conditional on a particular value of $y$, 
$f(x\mid y;G)$, which we refer to as inverse inferences, discussed further 
in the context of the data example of Section \ref{sec:ozone}.

\subsection{Prior specification}

We discuss two approaches to hyperprior specification considering the limiting 
case of the model as $\alpha\rightarrow 0^{+}$, which corresponds 
to a single mixture component. 
Both approaches use an approximate range and center of $x$, say $r^{x}$ and $c^{x}$, both vectors 
of length $p$, with the objective being to center and scale the mixture kernel 
appropriately using only a small amount of prior information. 
Under the assumption of a single mixture component, the marginal moments are 
given by $\mathrm{E}((z,x)^t)=a_m$, and $\mathrm{cov}((z,x)^t)=$ 
$\mathrm{E}(\Sigma)  + B_{m} + (a_{V}-p-2)^{-1} B_{V}$. We therefore set 
$a_m=(0,c^{x})$, and let $B_{m}=$ $0.5 \mathrm{diag}(1,(r^{x}_{1}/4)^2,...,(r^{x}_{p}/4)^2)$,
using $c^{x}_{j}$ and $(r^{x}_{j}/4)^2$ as proxies for the marginal mean and variance of 
$x_{j}$, for $j=1,...,p$.
We set $a_{V}=p+3$, which yields a dispersed prior for $V$ albeit with finite prior 
expectation, and determine $B_{V}$ such that $(a_{V}-p-2)^{-1} B_{V} = B_{m}$. 
Next, we must determine values for the prior hyperparameters associated with 
$\tilde{\beta}$ and the $\delta_{i}$, and this is where the two approaches differ. 

%
%

The first approach uses prior simulation to induce approximately uniform$(-1,1)$ 
priors on all correlations of the mixture kernel covariance matrix, while appropriately 
centering the variances. Note that the number of correlations grows at a rate of $O(p^2)$, 
making this approach practically feasible only for a small number of covariates. In particular, 
with a single covariate the kernel covariance matrix comprises correlation, $\rho=$
$-\tilde{\beta} (\tilde{\beta}^{2}+\delta)^{-1/2}$, and variance, $\sigma^{2}=$
$\tilde{\beta}^{2}+\delta$. Here, $\tilde{\beta}$ and $\delta$ are scalar parameters 
with $G_{0}$ components $\mathrm{N}({\theta},c)$ and $\mathrm{IG}(\nu,s)$, respectively, 
and the hyperpriors are: $\theta \sim \mathrm{N}(a_{\theta},b_{\theta})$, $c \sim \mathrm{IG}(a_c,b_c)$, and 
$s \sim \mathrm{gamma}(a_{s},b_{s})$. 
%
%
We set $\mathrm{E}(\tilde{\beta})= $ $a_{\theta}=0$, and build the 
specification for the other hyperparameters from 
$\mathrm{E}(\sigma^{2})=$ $b_{\theta} + b_{s}^{-1} (\nu-1)^{-1} a_{s} + (a_{c}-1)^{-1} b_{c}$.  
%
%
We first fix the shape parameters $\nu$, $a_{c}$ and $a_{s}$ to values that yield 
relatively large prior dispersion, for instance, $\nu=$ $a_{c}=2$ results in infinite 
prior variance for the inverse-gamma distributions.
Next, using $(r^{x}/4)^{2}$ as a proxy for $\mathrm{E}(\sigma^{2})$, we find constants 
$k_{1},k_{2},k_{3}$, where $k_{1} + k_{2} + k_{3} = 1$, such that $k_{1}(r^{x}/4)^{2} \approx b_{\theta}$, 
$k_{2}(r^{x}/4)^{2} \approx b_{s}^{-1} (\nu-1)^{-1} a_{s}$, and $k_{3}(r^{x}/4)^{2} \approx (a_{c}-1)^{-1} b_{c}$, 
while at the same time the induced prior on $\rho$ is approximately uniform on $(-1,1)$. 
Finally, with $k_{1},k_{2},k_{3}$ specified, $b_{\theta}$, $b_{s}$, and $b_{c}$ can be determined accordingly.

While this approach is attractive when a relatively noninformative prior is desired, it is
difficult to implement with a moderate to large number of covariates. An alternative 
strategy arises from studying the distribution which is implied for $(\beta,\Delta)$ if 
$\Sigma$ is inverse-Wishart distributed. Using properties of partitioned Wishart and 
inverse-Wishart matrices \citep{box,eaton}, it can be shown that 
$\Sigma \sim \mathrm{IW}_{p+1}(v,T)$ implies inverse-gamma distributions for the 
$\delta_{i}$, and a normal distribution for $\tilde{\beta}$ given the $\delta_{i}$.
It is customary to specify noninformative priors on the inverse-Wishart scale, 
usually fixing the degrees of freedom parameter to a small value, and the inverse 
scale parameter to be a diagonal matrix. 
Here, we use the smallest possible integer value for $v$ that ensures a finite expectation 
for the $\mathrm{IW}_{p+1}(v,T)$ distribution, that is, $v=p+3$, and set $\mathrm{E}(\Sigma)=$
$T=$ $\mathrm{diag}(T_1,\dots,T_{p+1})=$ $\mathrm{diag}(1,(r^{x}_{1}/4)^2,...,(r^{x}_{p}/4)^2)$. 
Then, as shown in Appendix B, the distributions implied on $\delta_{i}$, for $i=2,\dots,{p+1}$, 
are $\mathrm{IG}(0.5(v+i-(p+1)),0.5 T_{i})$. Hence, we let $\nu_i=$ $0.5 (v+i-(p+1))$, 
and $\mathrm{E}(s_i)=$ $0.5 T_{i}$; for the data examples of Section 3, we worked with 
exponential priors for the $s_{i}$ resulting in $b_{s_i}=2/T_{i}$.
Moreover, the $\mathrm{IW}_{p+1}(v,T)$ distribution implies a normal distribution for the 
$i$-th row of matrix $\beta$, given $\delta_{i}$; see Appendix B. This can be translated 
into a distribution for $\tilde{\beta}$ conditionally on the $\delta_{i}$, specifically, a normal 
distribution with zero mean vector and covariance matrix $\mathrm{BD}(S_1,\dots,S_p)$, which denotes 
a block diagonal matrix with elements $S_i=$ $\delta_{i+1} \mathrm{diag}(T_{1}^{-1},\dots,T_{i}^{-1})$,
for $i=1,...,p$. Now, after marginalizing out $\theta$, the $G_{0}$ prior component for $\tilde{\beta}$
becomes $\mathrm{N}_{q}(a_{\theta},B_{\theta}+C)$. We therefore specify $a_{\theta}$ to be equal to 
the zero mean vector, and since we have a further prior on $C$, and $S_i$ is a function of $\delta_{i+1}$, 
we set $B_{\theta}+\mathrm{E}(C)=$ $\mathrm{BD}(\hat{S}_1,\dots,\hat{S}_p)$, where $\hat{S}_i$ is a 
proxy for $S_{i}$ obtained by replacing $\delta_{i+1}$ with its marginal prior mean.
%
%
Finally, $B_{\theta}$ and $\mathrm{E}(C)$ can be specified to be equal to each other or 
assigned different portions of $\mathrm{BD}(\hat{S}_1,\dots,\hat{S}_p)$.

\section{Data Illustrations}

\subsection{Ozone data}
\label{sec:ozone}
Ozone is a gas which has detrimental consequences when it occurs near the Earth's surface. 
Ground-level ozone is a harmful pollutant, making up most of the smog which is visible 
in the sky over large cities. Because of the effects ozone has on the environment and our 
health, its concentration is monitored by 
environmental agencies. Rather than recording the actual concentration, presence or absence 
of an exceedance over a given ozone concentration threshold may be measured, and the 
number of ozone exceedances in a particular area is of interest. 

We work with data set {\tt ozone} from the ``ElemStatLearn'' R package. The data set
includes measurements of ozone concentration in parts per billion, wind speed in miles per hour, 
temperature in degrees Fahrenheit, and radiation in langleys, recorded over 111 days from 
May to September of 1973 in New York. 
To construct a binary ozone exceedance response, we define an exceedance as an ozone 
concentration which is larger than 70 parts per billion. Therefore, we can model the probability 
of an ozone exceedance as a function of wind speed, temperature, and radiation, using the 
DP mixture binary regression model. In addition, the modeling approach is evidently appropriate here, since it is natural to estimate
conditional relationships between the four environmental variables through modeling the 
stochastic mechanism for their joint distribution. 
We are not suggesting dichotomizing a continuous response in practice, but use 
this example to illustrate a practically relevant setting in which a binary response 
may arise as a discretized version of a continuous response. Moreover, the existence of the continuous ozone concentrations enables comparison 
of inferences from the binary regression model with a model based on the 
actual continuous responses.

Prior specification was performed using the first approach discussed in Section 2.3
that favors uniform priors for the correlations of the kernel covariance matrix. 
Although the corresponding priors were not all close to the uniform on $(-1,1)$
under the inverse-Wishart prior specification approach, both methods resulted in prior mean 
estimates for $\mathrm{Pr}(y=1\mid x_j)$ that were, for each of the three random covariates,
constant around $0.5$, with $90\%$ interval bands that essentially span the unit interval. 
All posterior inference results discussed below were robust to the prior choice.

\begin{figure}[t]
\begin{tabular}{ccc}
\includegraphics[height=1.9in,width=0.31\textwidth]{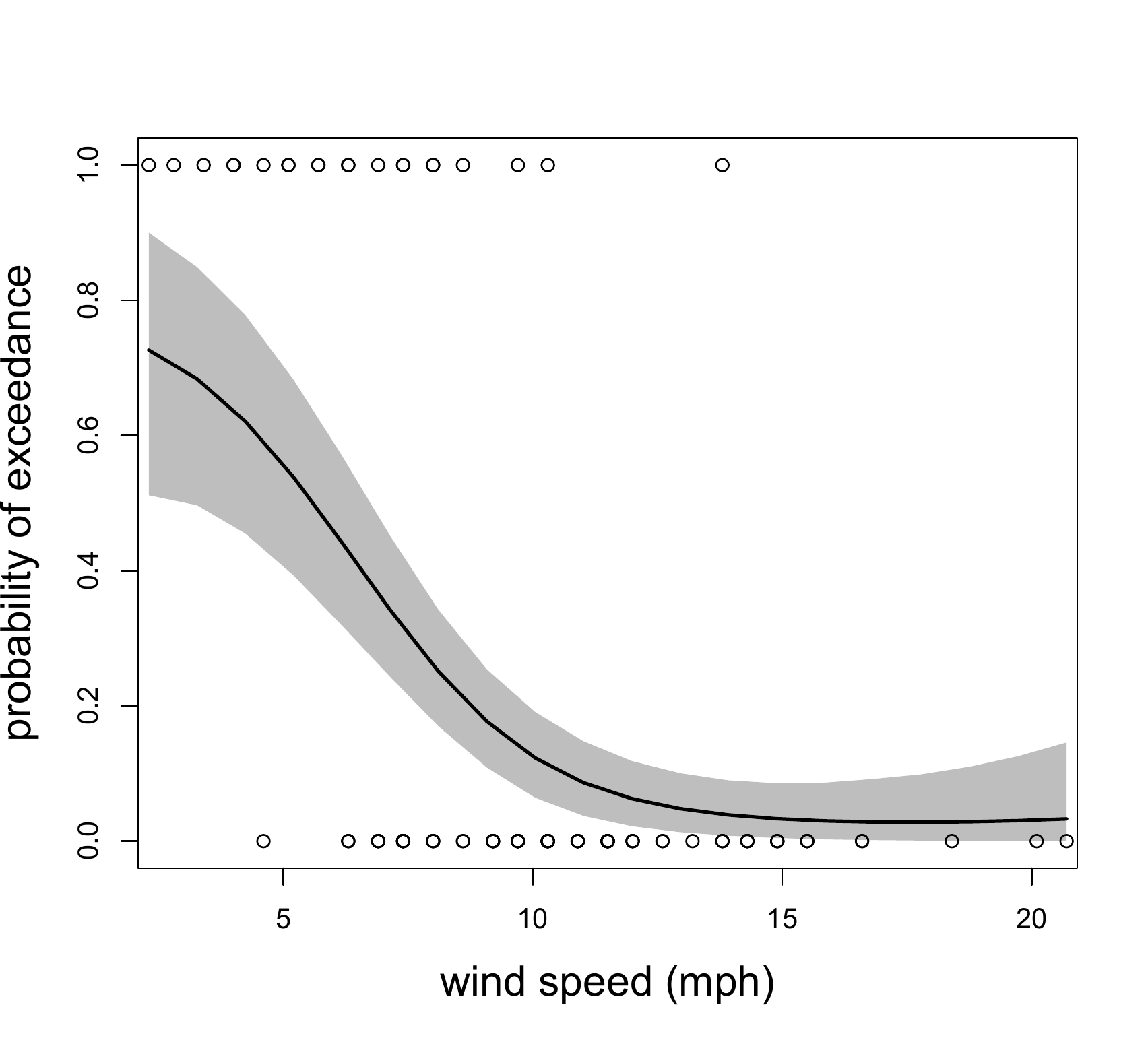} &
\includegraphics[height=1.9in,width=0.31\textwidth]{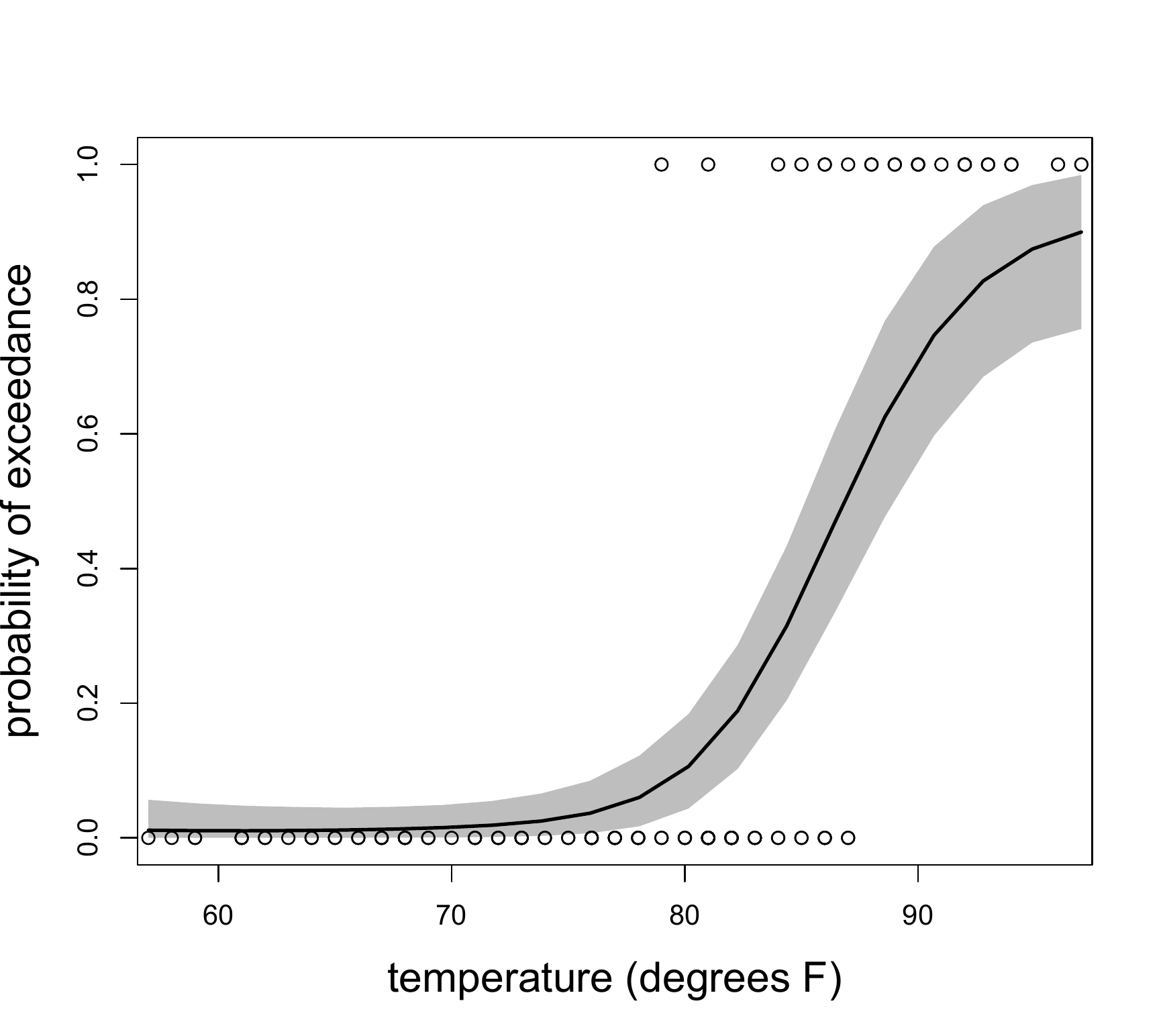} &
\includegraphics[height=1.9in,width=0.31\textwidth]{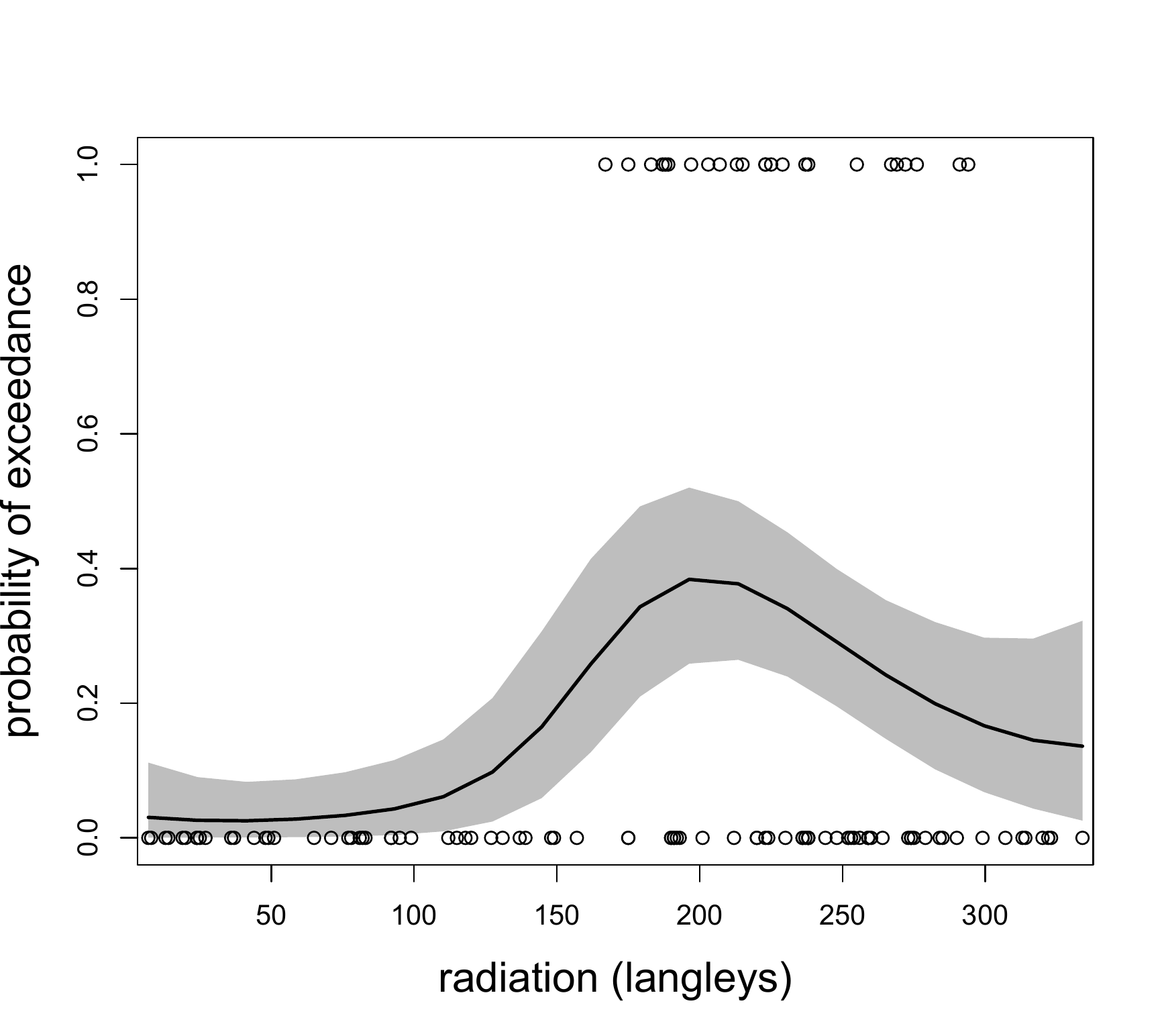} \\
\includegraphics[height=1.9in,width=0.31\textwidth]{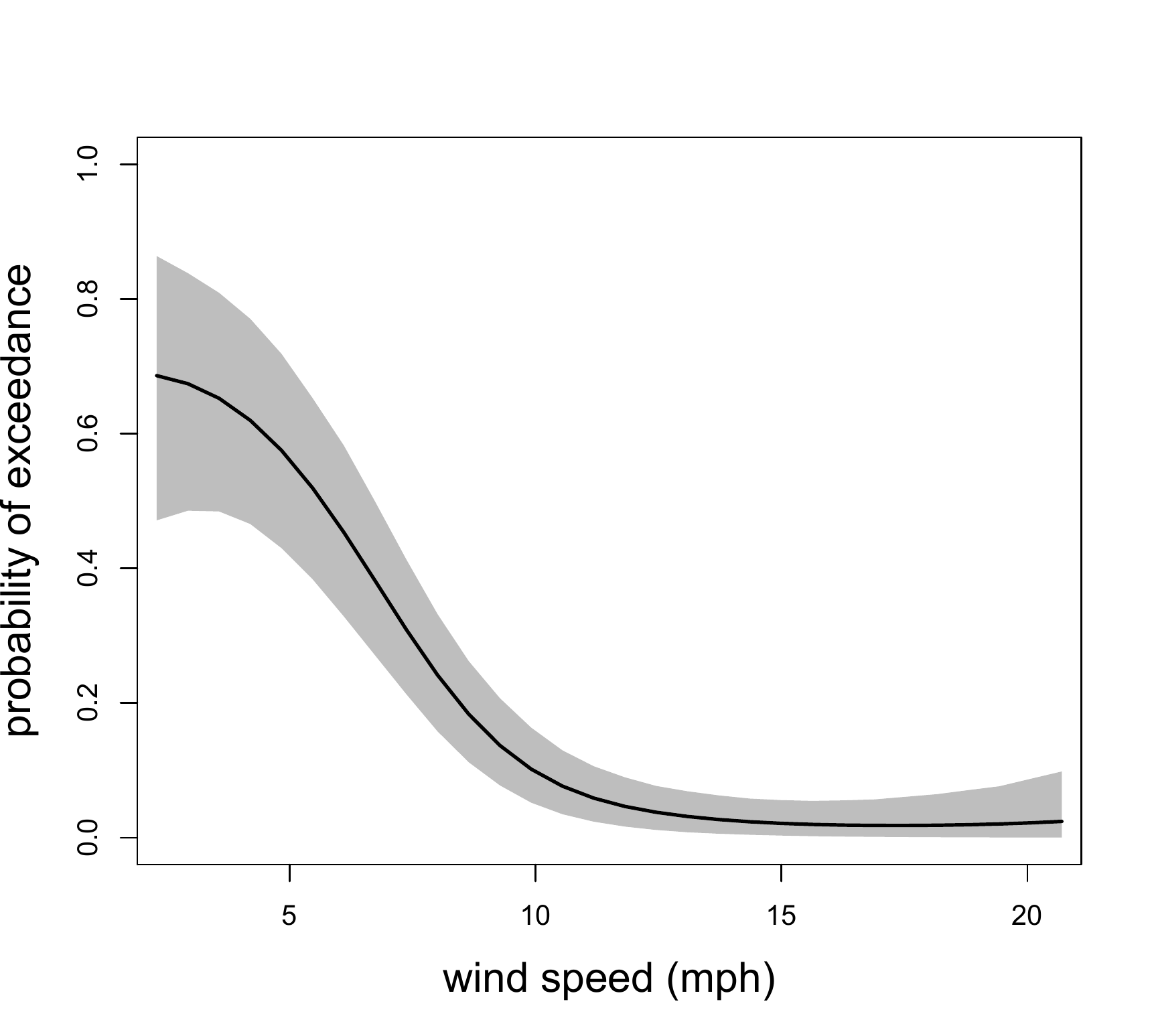} &
\includegraphics[height=1.9in,width=0.31\textwidth]{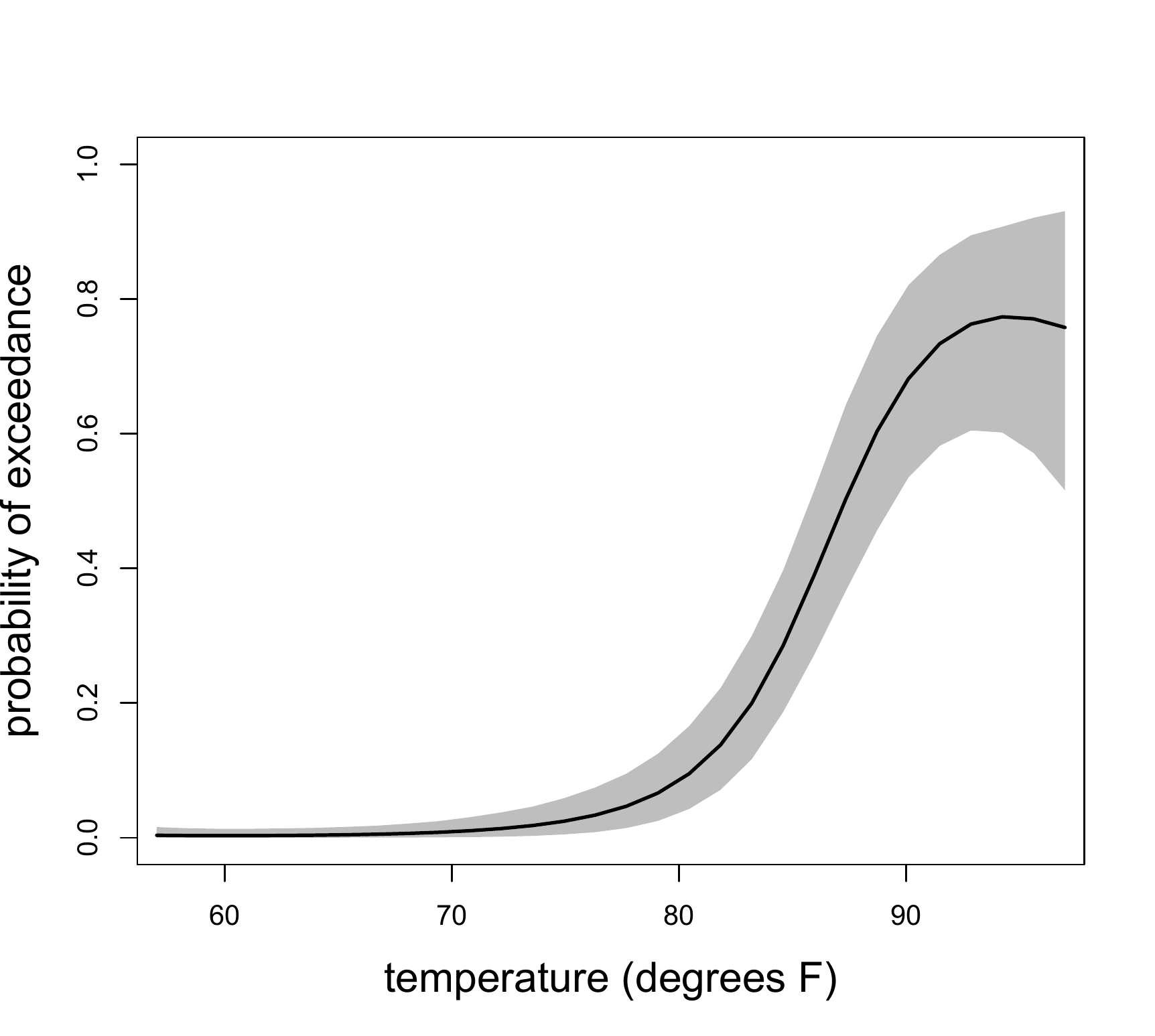} &
\includegraphics[height=1.9in,width=0.31\textwidth]{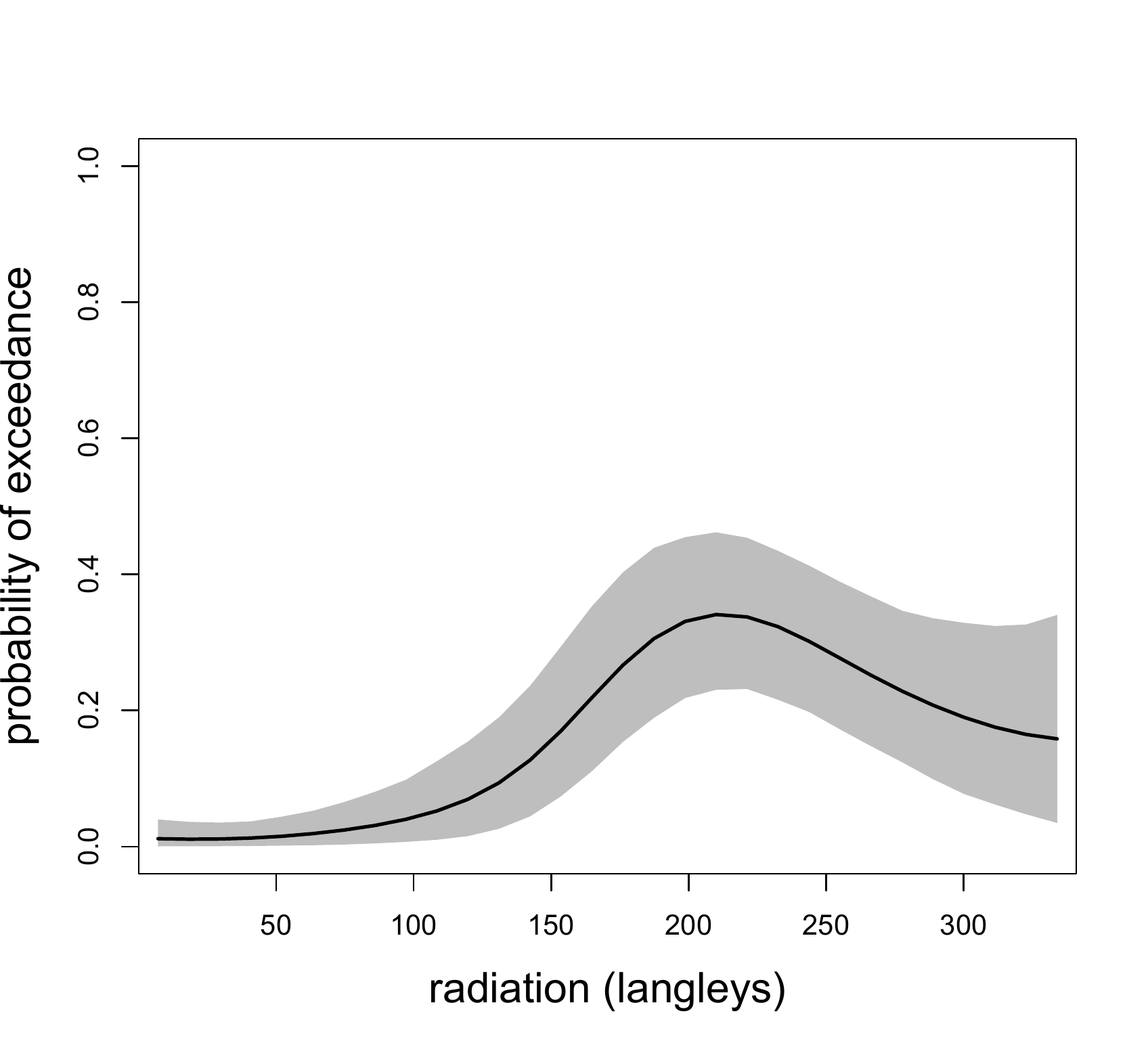}
\end{tabular}
\caption{Ozone data.
Posterior mean (solid line) and $90\%$ uncertainty bands (in gray) 
for probability of exceedance versus wind speed (left panels), temperature (middle panels), 
and radiation (right panels). The top row plots results under the binary regression
model, including the binary response data in each panel. The bottom row shows
results under the density estimation model. Refer to Section 3.1 for further details.}
\label{ozone-prob}
\end{figure}

\begin{figure}[t]
\begin{tabular}{cc}
\includegraphics[height=2.75in,width=0.49\textwidth,angle=270]{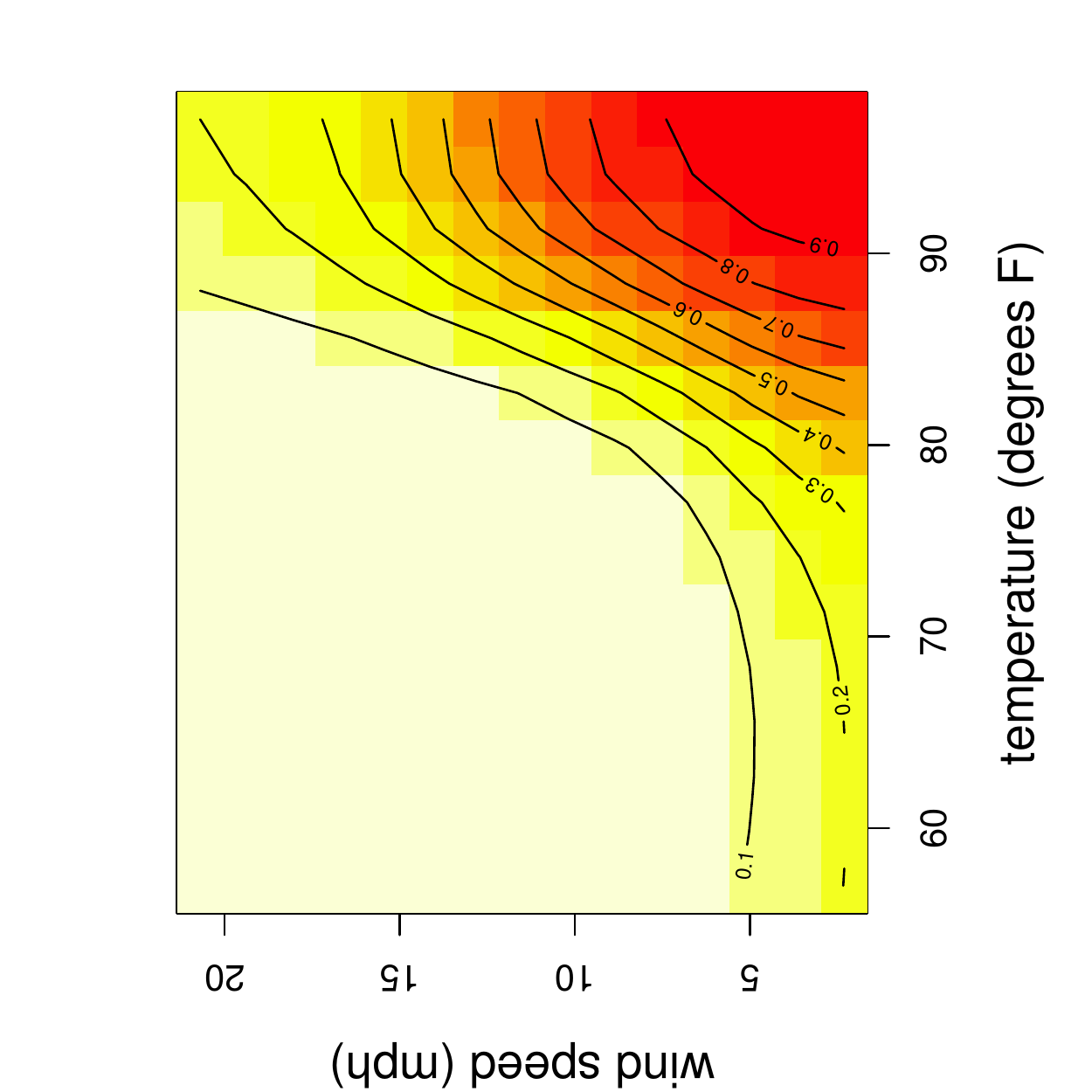}
\includegraphics[height=2.75in,width=0.49\textwidth,angle=270]{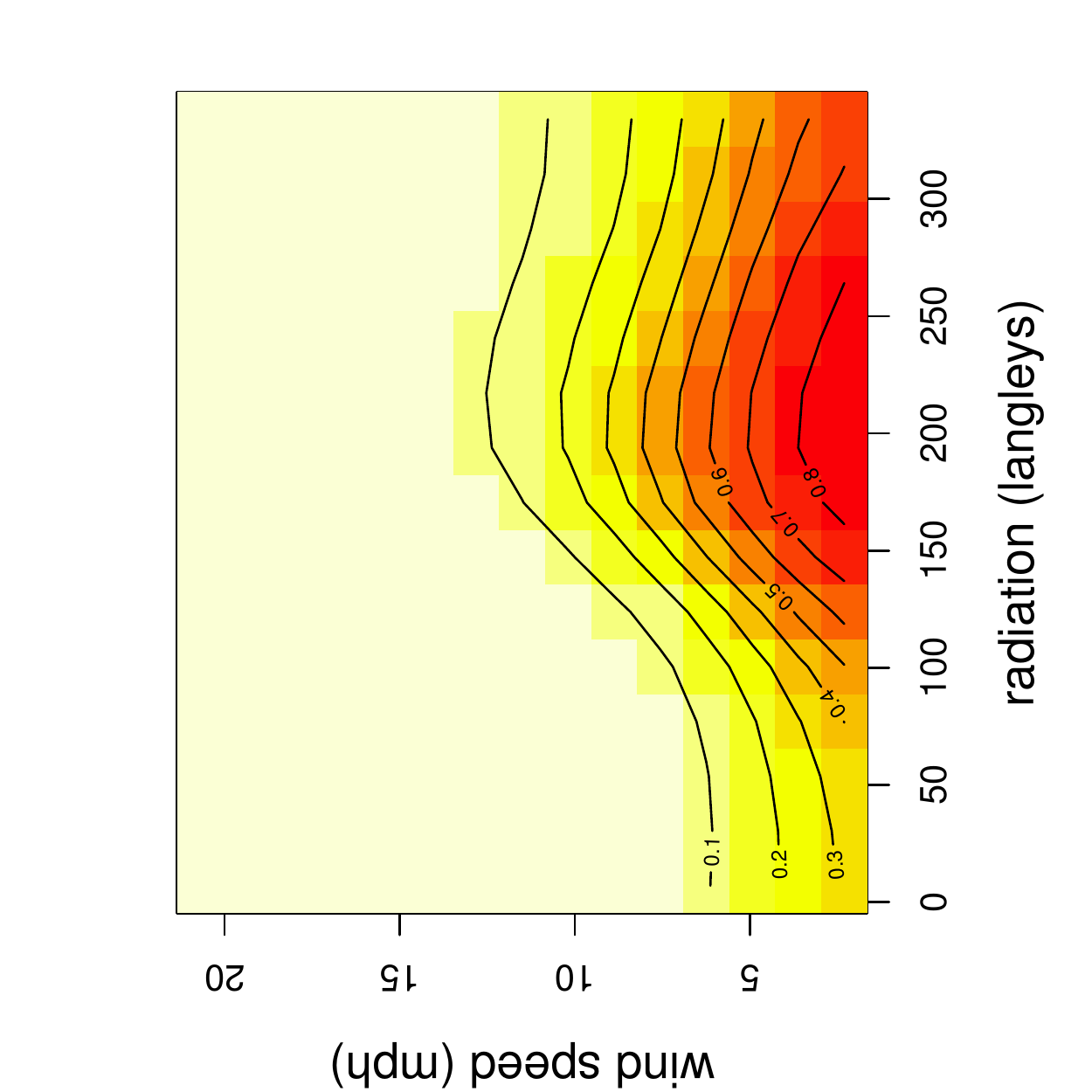}
\end{tabular}
\caption{Ozone data.
Posterior mean surface for probability of exceedance versus 
temperature and wind speed (left panel), and radiation and wind speed (right panel). 
Probabilities ranging from 0 to 1 are indicated by a spectrum of colors from white to red.}
\label{ozone-bivariate-prob}
\end{figure}

The marginal binary response curves for the probability of exceedance as a function of wind speed, 
temperature, and radiation, are shown in the top row of Figure \ref{ozone-prob}. There is a 
decreasing trend in probability as wind speed increases, with the probability 
being essentially 0 when wind speed is greater than 15 mph. 
The opposite trend is observed with temperature, as the probability of exceedance is near 0 
when temperature is less than 75 degrees, and above 0.8 when temperature exceeds 90 degrees. 
A non-monotonic unimodal response curve is obtained as a function of radiation, 
with peak probability occurring at moderate values of radiation, and declining with higher and 
lower values. Bivariate surfaces indicating probability of exceedance as a function of temperature 
and wind speed, as well as radiation and wind speed, are shown in Figure \ref{ozone-bivariate-prob}. 
An attractive feature of the joint modeling approach is that interactions and dependence 
between covariates are naturally accounted for, without the need to make simplifying 
assumptions, such as additivity in covariate effects, or to accommodate interactions 
with additional terms.

For this illustrative data example, the continuous ozone concentration responses are also 
available. We can therefore compare the binary regression model 
inferences for $\mathrm{Pr}(y=1\mid x_j)$ with the ones for $\mathrm{Pr}(z>70\mid x_j)$,
under the corresponding density estimation model -- a DP mixture based on a 
four-dimensional normal kernel with unrestricted covariance matrix -- applied to the original 
data set $\{(z_{i},x_{i}):i=1,...,111\}$. Results are shown in the bottom row of Figure \ref{ozone-prob}, 
based on a prior choice for the density estimation model that induces prior estimates for 
the $\mathrm{Pr}(z>70\mid x_j)$ curves that are similarly diffuse to the ones for $\mathrm{Pr}(y=1\mid x_j)$.
Save for some differences in the uncertainty bands, the density estimation model reveals 
similar trends for the regression functions to the ones uncovered by the binary regression model.

\begin{figure}
\begin{tabular}{ccc}
\includegraphics[height=2in,width=0.31\textwidth]{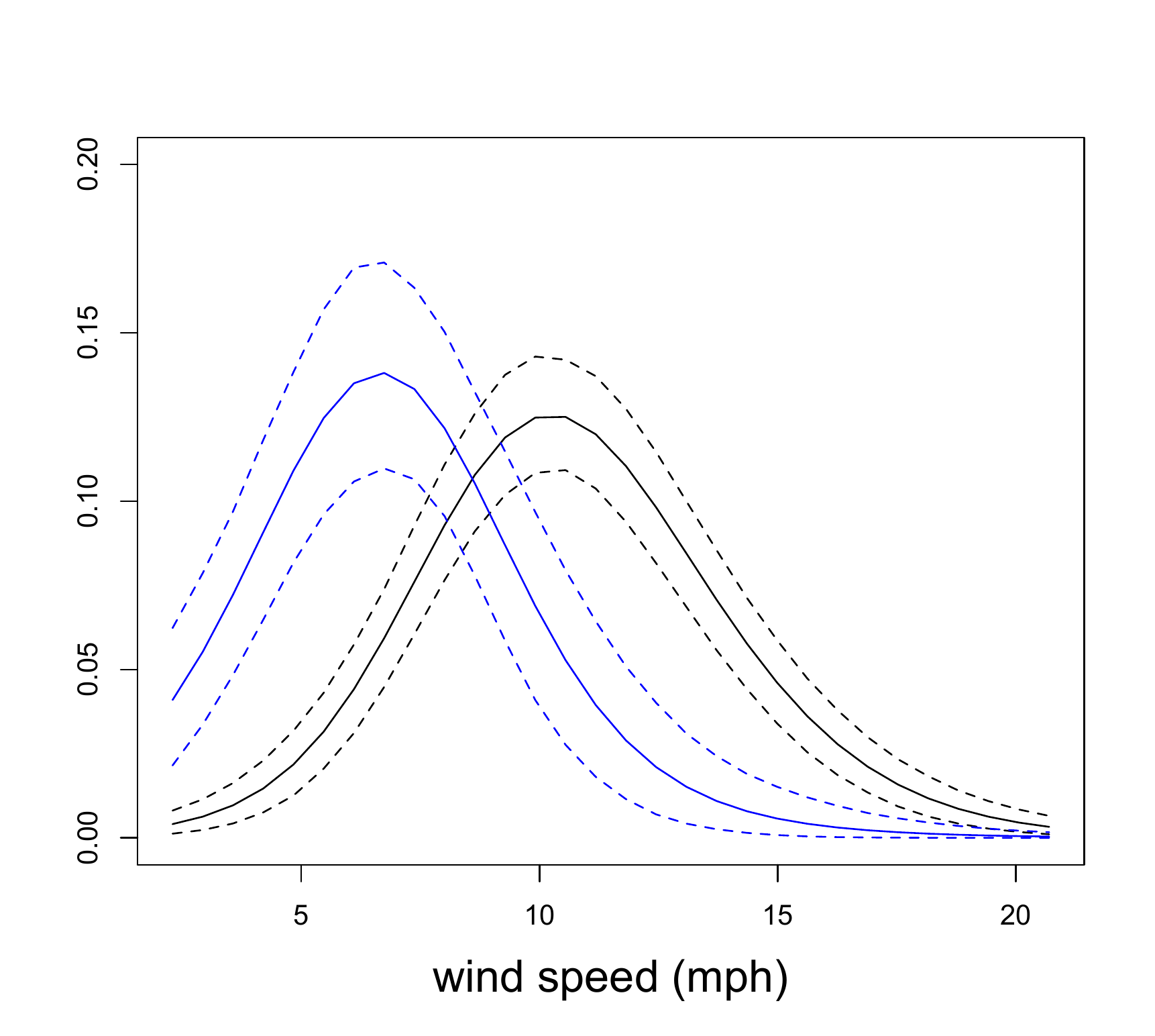} &
\includegraphics[height=2in,width=0.31\textwidth]{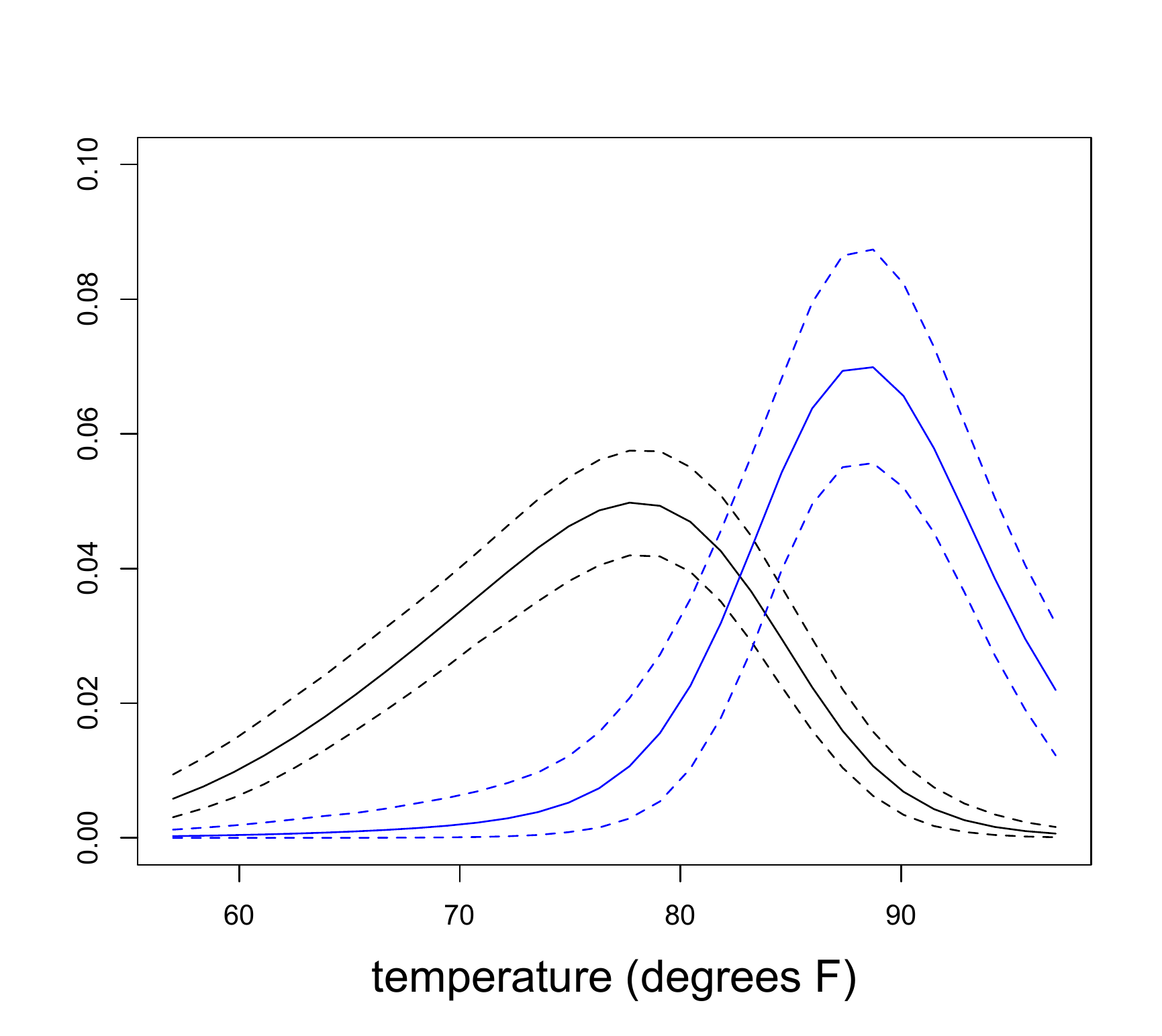} &
\includegraphics[height=2in,width=0.31\textwidth]{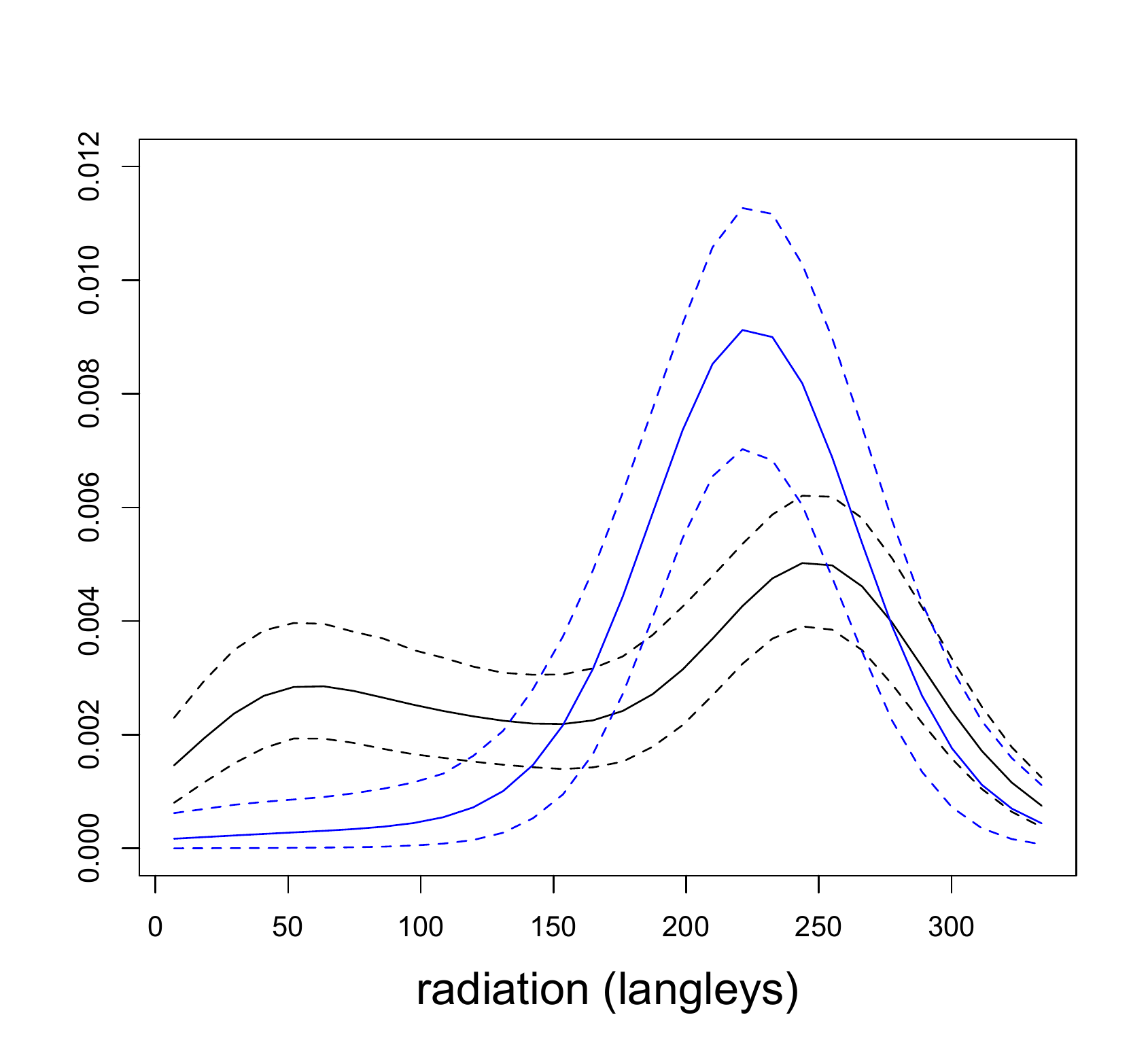} 
\end{tabular}
\caption{Ozone data.
Posterior mean estimates (solid lines) and $90\%$ uncertainty bands (dashed lines) 
for the density of wind speed (left panel), temperature (middle panel), 
and radiation (right panel), given an ozone concentration exceedance (blue) 
and non-exceedance (black).}
\label{ozone-densities}
\end{figure}

As another appealing consequence of estimating the joint response-covariate distribution, 
we can obtain inference for the distribution of covariates conditional on a particular value 
of $y$. These inverse inferences may be of interest in many settings, as they indicate how the 
covariate distribution differs
given a positive versus a negative binary response.
Such inferences are not possible under a model directly for the conditional response 
distribution (with the implicit assumption of fixed covariates). 
Figure \ref{ozone-densities} shows estimates for the density 
of each covariate conditional on the binary exceedance response, 
$f(x_j \mid y=1)$ and $f(x_j \mid y=0)$, for $j=1,2,3$. 
Note that when an exceedance occurs, temperature is generally higher and 
wind speed lower. In addition, the conditional densities associated with an 
exceedance have smaller dispersion than those associated with a 
non-exceedance, indicating that a smaller range of covariate values 
are supported when an exceedance occurs.

Recall from Section 2.1 that if we make the simplifying assumption ${\Sigma}^{zx}={0}$ for the 
covariance matrix of the kernel in $f(z,x;G)$, we obtain a kernel for $f(y,x;G)$ that comprises
independent components $\mathrm{N}_{p}(x;\mu^{x},{\Sigma}^{xx})$ and $\mathrm{Bern}(y;\Phi(\mu^{z}))$.
The implied conditional regression function is again a weighted sum of probabilities with the same 
covariate-dependent weights as the proposed model, but probabilities which are not functions of $x$; 
the probability $\pi_{l}(x)$ in expression (\ref{eqn:piweights}) reduces to $\pi_l=$ $\Phi(\mu_{l}^{z})$. 
Mixtures of this product-kernel form have been previously proposed in the literature; see, for 
instance, \citet{dunson}.

We fitted the simpler product-kernel model to the ozone data, using hyperpriors that induce similarly 
diffuse prior estimates for the regression functions with the general binary regression model. Differences 
in the response probabilities produced by the product-kernel mixture model (not shown here)
tend to occur at peaks or low points of the curves in Figure \ref{ozone-prob}.
In general, the product-kernel model underestimates the probability surface or curve 
when it takes a high value, and overestimates regions of low probability. In addition, the uncertainty 
bands from the product-kernel model are generally wider than those produced by the proposed model.

%
%
For a more formal comparison, we use the posterior predictive loss criterion of \citet{gelfand}. 
The criterion favors the model $m$ that minimizes the predictive loss measure $D_{k}(m)=$ 
$P(m)+\{k/(k+1)\}G(m)$, with penalty term $P(m)=$ $\sum_{i=1}^{n}\mathrm{var}^{(m)}(y_{new,i}\mid\mathrm{data})$, 
and goodness of fit term $G(m)=$ $\sum_{i=1}^{n}\{y_{i} - \mathrm{E}^{(m)}(y_{new,i}\mid\mathrm{data})\}^2$. Here, 
$\mathrm{E}^{(m)}(y_{new,i}\mid\mathrm{data})$ is the mean under model $m$ of the posterior predictive distribution 
for replicated response $y_{new,i}$ with corresponding covariate value $x_i$. The variance is similarly defined.  
Details involving expressions contributing to $D_{k}(m)$ for each model are given in Appendix C, but note that 
computations are based on the conditional posterior predictive distribution of $y$ given $x$. 
The penalty term under the product-kernel model is $10.17$, while it is $7.95$ under the proposed model, 
and the goodness of fit terms are $4.17$ and $4.08$, respectively. 
%
Hence, regardless of the choice for constant $k$, the criterion favors the general DP mixture
binary regression model.
%
%

\subsection{Estimating natural selection functions in song sparrows}

In addition to enabling more general modeling of binary regression relationships, the 
latent variables may be practically relevant in specific applications. 
Often, we may only observe whether or not some event occurred,
although there exists an underlying continuous response which drives the binary observation. 
The ozone data was used to illustrate an environmental application for which the 
latent continuous responses are actually present. In applications in biology, the latent 
response may represent maturity, which is recorded on a discretized scale, or an 
unobservable trait or measure of health. In general, the continuous responses may be 
latent either because they are actually unobservable, or as consequence of recording 
taking place on a discretized scale. As an example of the former scenario, consider a 
binary response which represents survival. While we only observe survival on a binary scale, 
it is meaningful to conceptualize an underlying process which drives survival.
Quantifying the probability of survival as a function of phenotypic traits is of great interest 
in evolutionary biology \citep{lande, schluter, janzen}.  
Survival can be thought of as a measure of fitness, and the fitness surface describes the 
relationship between phenotypic traits and fitness.  
The proposed methodology is particularly well-suited for this area of application,
as it allows flexible inference for the shape of the fitness surface and for the distribution
of population traits under a joint modeling framework that incorporates the scientifically
relevant latent fitness responses. 

As an illustration, we consider a standard data set from the relevant
literature that records overwinter mortality along with six
morphological traits in a population of $145$ female song sparrows
\citep{schlutersmith}. The traits measured consist of weight,
wing length, tarsus length, beak length, beak depth, and beak width. 
Our initial analysis included four traits -- weight, wing length,
tarsus length, and beak length -- as beak width and depth are highly 
discretized, correlated with beak length, and did not appear to be 
associated with a trend in survival. This analysis revealed tarsus
length and beak length to be the main targets of selection, which is 
consistent with the findings of \citep{schlutersmith}. A key 
objective in this example is to obtain inferences for functionals 
used to assess the strength and form of natural selection acting 
on phenotypic traits, and we thus focus on the two traits associated with survival. 

\begin{figure}[t]
\begin{tabular}{cc}
\includegraphics[height=2.75in,width=0.48\textwidth]{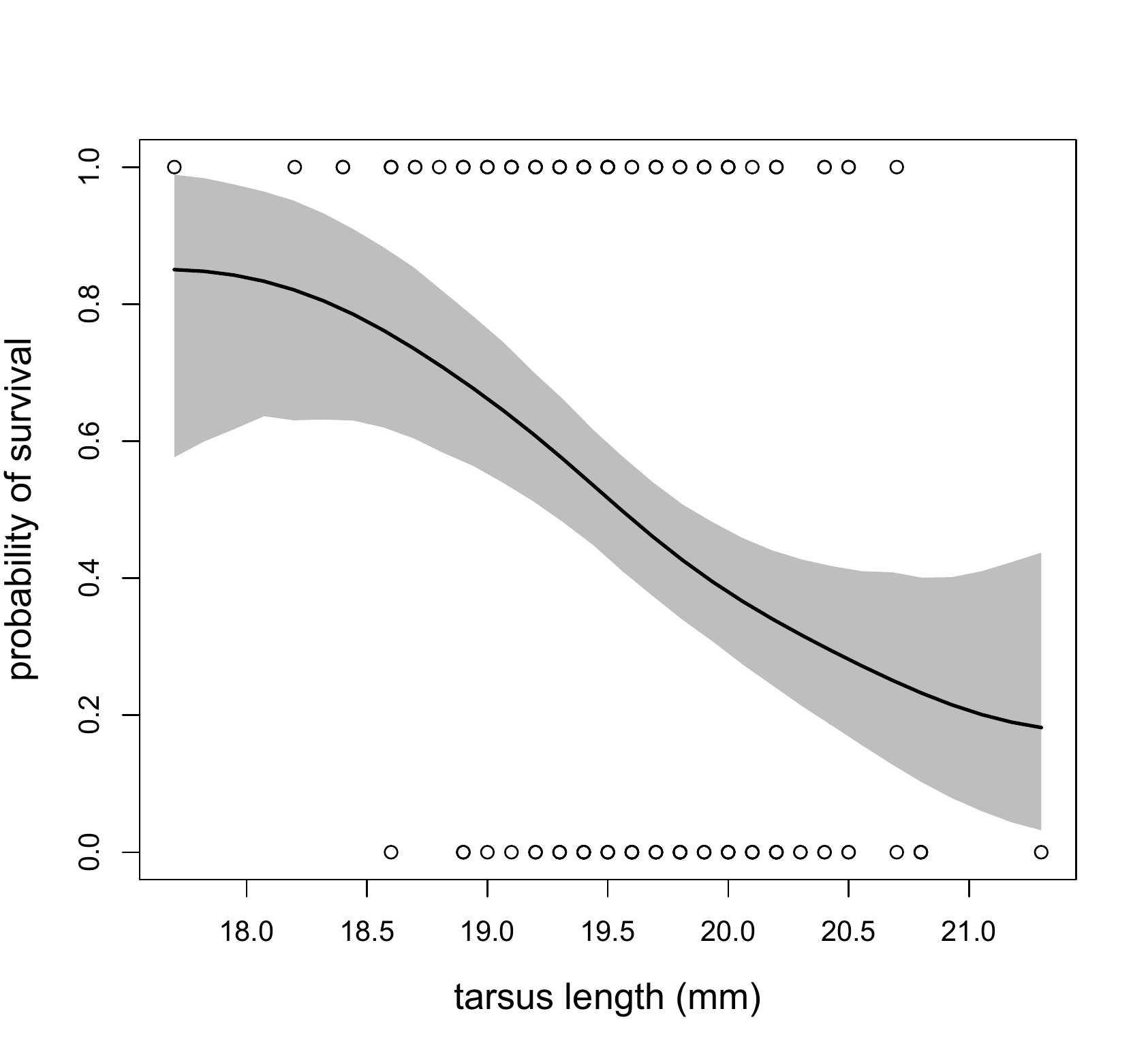}
\includegraphics[height=2.75in,width=0.48\textwidth]{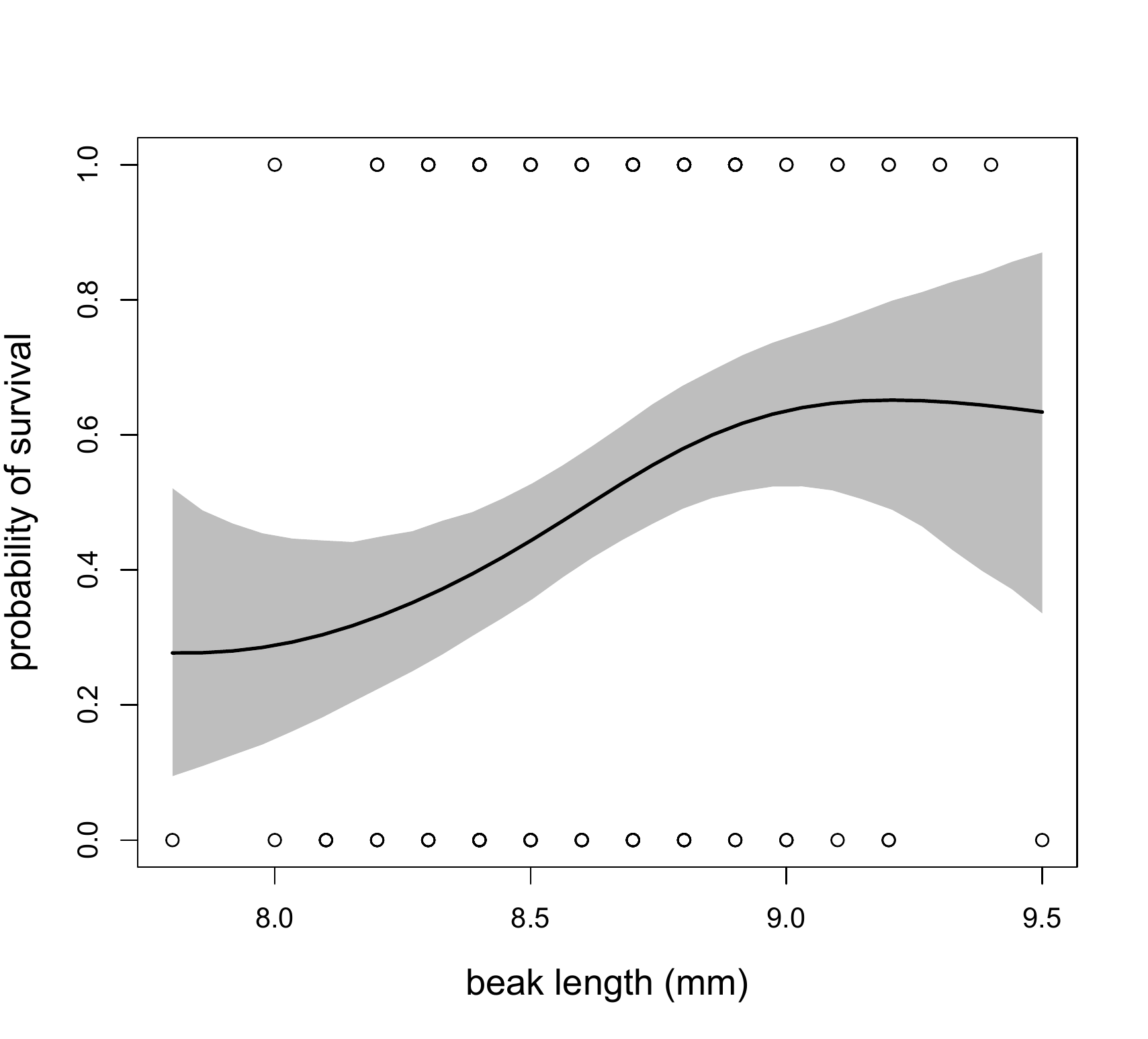}
\end{tabular}
\caption{Song sparrows data. Posterior mean (solid line) and $90\%$ uncertainty 
bands (in gray) for the probability of survival as a function of
tarsus length (left panel) and beak length (right panel). Plotted in each panel
are the corresponding observations.}
\label{selection_curves}
\end{figure}

\begin{figure}
\begin{tabular}{cc}
\includegraphics[height=2.75in,width=0.48\textwidth]{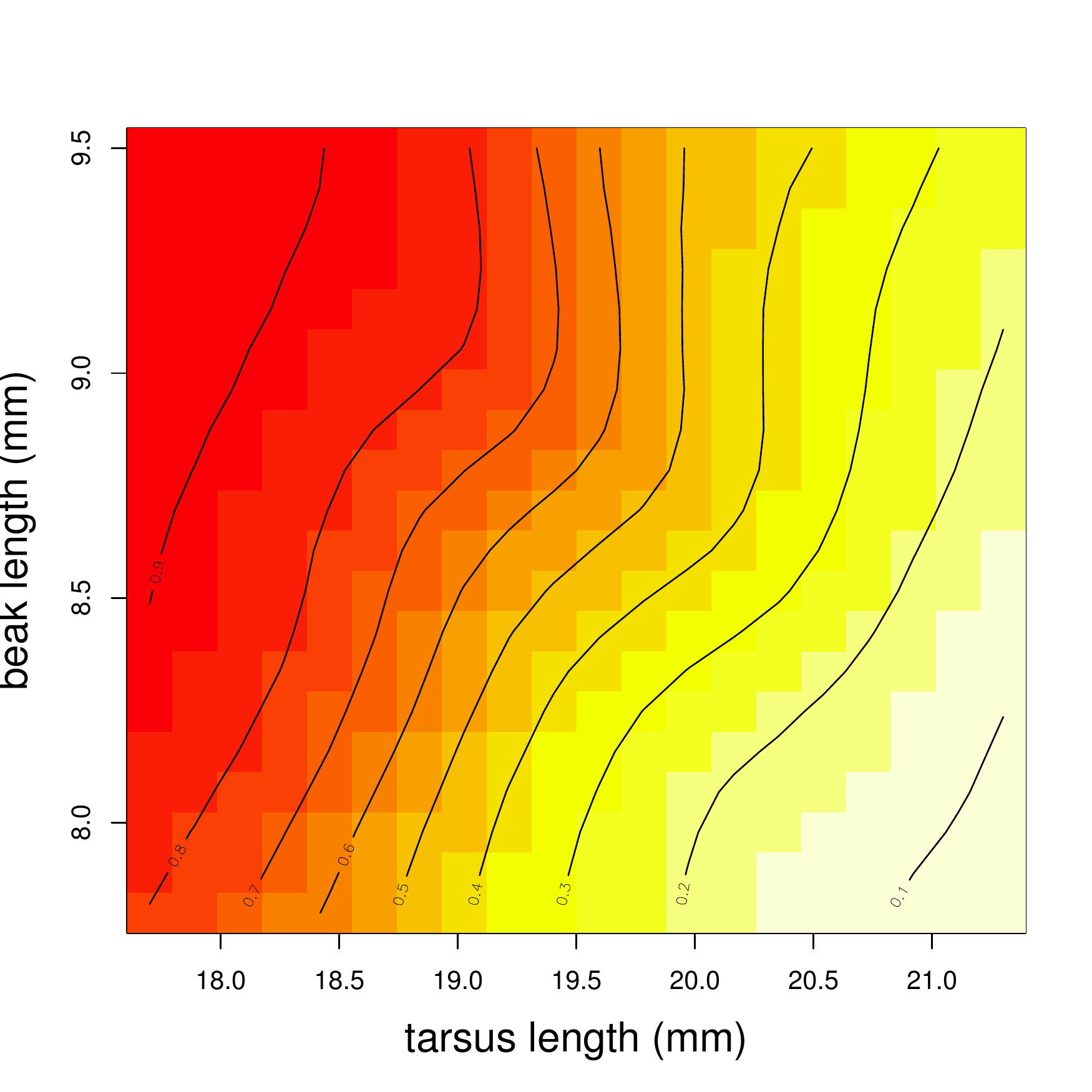}
\includegraphics[height=2.75in,width=0.48\textwidth]{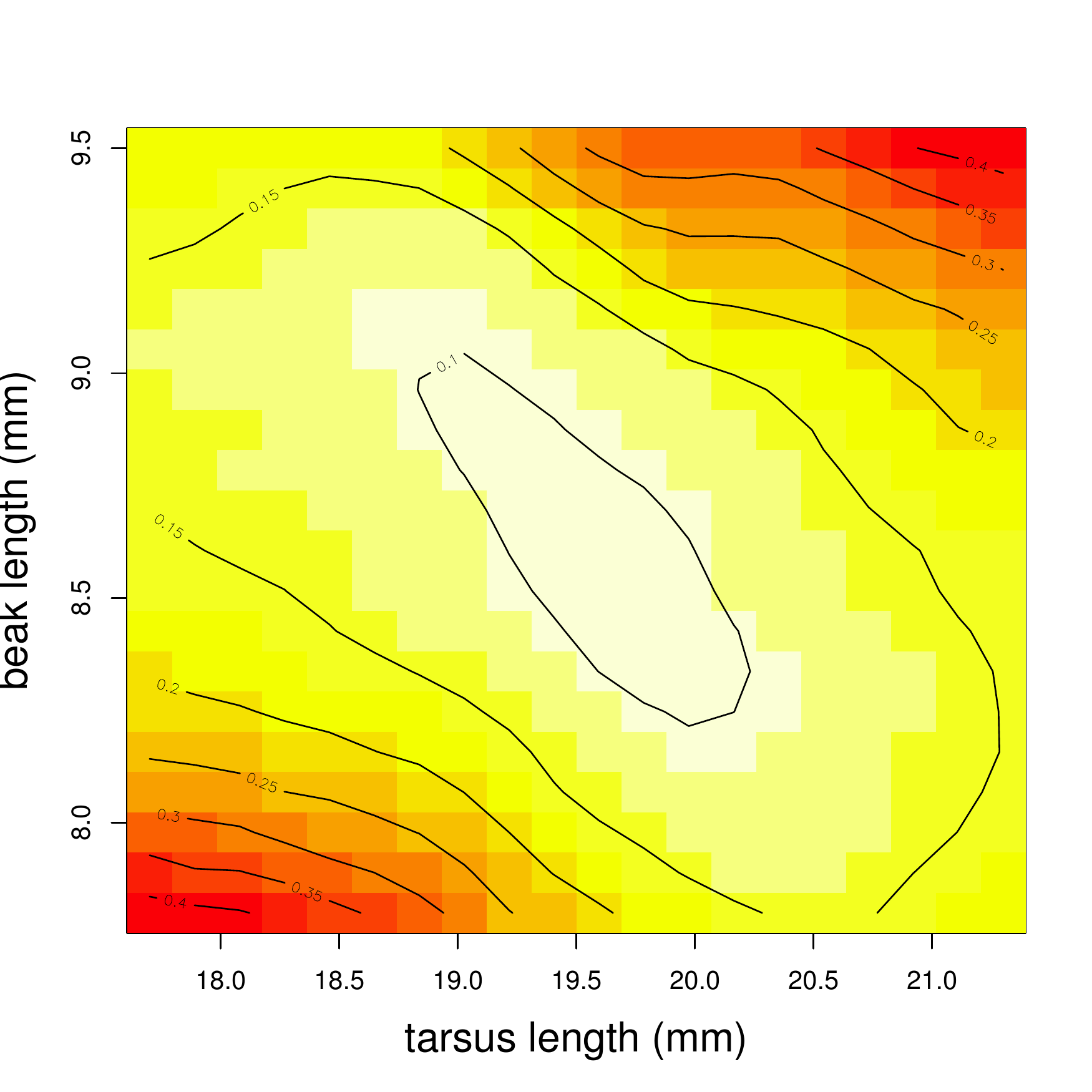}
\end{tabular}
\caption{Song sparrows data. Posterior median surface (left panel) and interquartile range
surface (right panel) for the probability of survival as a function of
tarsus length and beak length.}
\label{selection_surface}
\end{figure}

The model was applied with standardized covariates tarsus length
($x_1$) and beak length ($x_2$), measured in millimeters, using the
second approach to prior specification involving the inverse-Wishart
distribution. The estimated selection curves are shown in Figure
\ref{selection_curves}, revealing a strong decreasing trend in fitness
over tarsus length, in which a sparrow with tarsus length $20.55$
millimeters has a $10\%$ lower probability of surviving overwinter
than a sparrow with tarsus length just $0.5$ mm shorter. The 
opposite trend in fitness is present over beak length, as longer beaks 
are associated with higher probabilities of survival. 
The posterior median estimate for the probability of survival as a function of both 
traits (Figure \ref{selection_surface}, left panel) confirms that the combination 
of long beaks and short tarsi is optimal for fitness; importantly, it also indicates 
that a short tarsus provides the more significant contribution to higher probability
of survival. The corresponding posterior interquartile range estimate 
(Figure \ref{selection_surface}, right panel) depicts more uncertainty in the 
survival probability surface for sparrows having both a short beak and short 
tarsus, and those with both a long beak and long tarsus.

For each of the two traits, we estimated the standardized directional
selection differential, $\bar{x}_{j}^{*} - \bar{x}_{j}$, which provides
a measure of selection intensity representing the change in mean
value of a phenotype produced by selection (Lande and Arnold, 1983). 
Here, $\bar{x}_{j}=$ $\int x_{j} f(x_{j}) \mathrm{d}x_{j}$ is the mean value of 
phenotypic trait $x_j$ before selection, and $\bar{x}_{j}^{*}=$ 
$\int x_{j} f(x_{j} \mid y=1) \mathrm{d}x_{j}=$ 
$\{ \mathrm{Pr}(y=1) \}^{-1} \int x_{j} \mathrm{Pr}(y=1,x_j) \mathrm{d}x_j$
is the mean value after selection; the marginal probability $\mathrm{Pr}(y=1)$
is referred to as mean absolute fitness.
Under our model, $\bar{x}_{j}=$ $\sum_{l=1}^{N} p_{l} \mu_{l}^{x_j}$, 
the mean absolute fitness is given by $\sum_{l=1}^{N} p_{l} \Phi(\mu_{l}^{z})$,
and $\int x_{j} \mathrm{Pr}(y=1,x_{j};G_N) \mathrm{d}x_j$ is approximated
with a Riemann sum.
The posterior mean estimate for the standardized selection differential for tarsus
length was $-0.31$, with a $90\%$ posterior credible interval of
$(-0.46,-0.18)$.  For beak length, the posterior mean and $90\%$ credible
interval for the standardized selection differential were $0.22$ and
$(0.09,0.36)$. Note that these intervals do not contain zero. Combined
with the estimated regression curves, these results give strong
evidence that directional selection is acting on tarsus length and
beak length, favoring sparrows with long beaks and short tarsi. 
%
%

The average gradient of the selection surface, weighted by the
phenotype distribution, is given under our model by the vector 
\[
\left(\int \frac{\partial \mathrm{Pr}(y=1\mid x;G_N)}{\partial x_1}f(x;G_N) \mathrm{d}x, 
\int\frac{\partial \mathrm{Pr}(y=1\mid x;G_N)}{\partial x_2}f(x;G_N) \mathrm{d}x\right)^{t}.
\] 
Under a linear regression structure with a multivariate normal distribution 
for the phenotypic traits, the selection gradient is equivalent to the
vector of linear regression slopes \citep{lande}. 
\citet{janzen}  do not incorporate in their approach a
distributional assumption for $f(x)$, and approximate the $j$-th 
selection gradient by $n^{-1} \sum_{i=1}^{n} \partial \mathrm{Pr}(y=1\mid x)/\partial
x_j\mid_{x=x_i}$. Our joint mixture modeling approach avoids the
assumption of normality for the phenotypic distribution, as well as the
need to estimate the integral by assuming the sample represents the
population distribution. The integrand of the $j$-th component of the 
selection gradient vector can be written as 
$\{ \partial\mathrm{Pr}(y=1,x;G_N)/\partial x_j \} -
\{ \mathrm{Pr}(y=1\mid x;G_N)\partial f(x;G_N)/\partial x_j \}$, for $j=1,2$.
We omit the specific expressions for each of these two terms, but note 
that both are analytically available as a consequence of the mixture of normals
representation for $f(z,x;G_N)$. 
Finally, the average gradient of the relative selection surface, also 
referred to as the directional selection gradient by \citet{lande}, 
is obtained by dividing each element of the selection gradient vector
by mean absolute fitness.
We obtained posterior mean estimates of $-0.27$ and $0.18$, with corresponding 
$90\%$ credible intervals of $(-0.40,-0.14)$ and $(0.06,0.31)$, for the directional 
selection gradient associated with tarsus length and beak length, respectively.

The presence of stabilizing or disruptive selection can be explored by
considering the change in the phenotypic variance-covariance matrix
due to selection, that is, the change from the pre-selection
covariance matrix $P$, with elements 
$\int (x_{1}-\bar{x}_{1},x_{2}-\bar{x}_{2})^{t} (x_{1}-\bar{x}_{1}, x_{2}-\bar{x}_{2})
f(x) \mathrm{d}x$, to the post-selection covariance matrix $P^{*}$, with elements
$\int (x_{1}-\bar{x}_{1}^{*},x_{2}-\bar{x}_{2}^{*})^{t} (x_{1}-\bar{x}_{1}^{*}, x_{2}-\bar{x}_{2}^{*})
f(x\mid y=1) \mathrm{d}x$.
The stabilizing selection differential matrix is given by $P^*-P$ +
$(\bar{x}_{1}^{*}-\bar{x}_{1},\bar{x}_{2}^{*}-\bar{x}_{2})^{t} 
(\bar{x}_{1}^{*}-\bar{x}_{1},\bar{x}_{2}^{*}-\bar{x}_{2})$ (Lande and Arnold, 1983), 
where negative values for a particular trait indicate the presence of
stabilizing selection, while positive values indicate disruptive
selection. The posterior mean for the matrix element corresponding to
tarsus length is $0.038$, that for beak length is $-0.020$, and the
off-diagonal element has a posterior mean of $-0.018$. The $90\%$ 
posterior credible intervals for each element of the matrix all include zero,
indicating lack of significant evidence for stabilizing or disruptive
selection acting on either trait.

\begin{figure}[t]
\begin{tabular}{ccc}
\includegraphics[height=2.4in,width=0.95\textwidth]{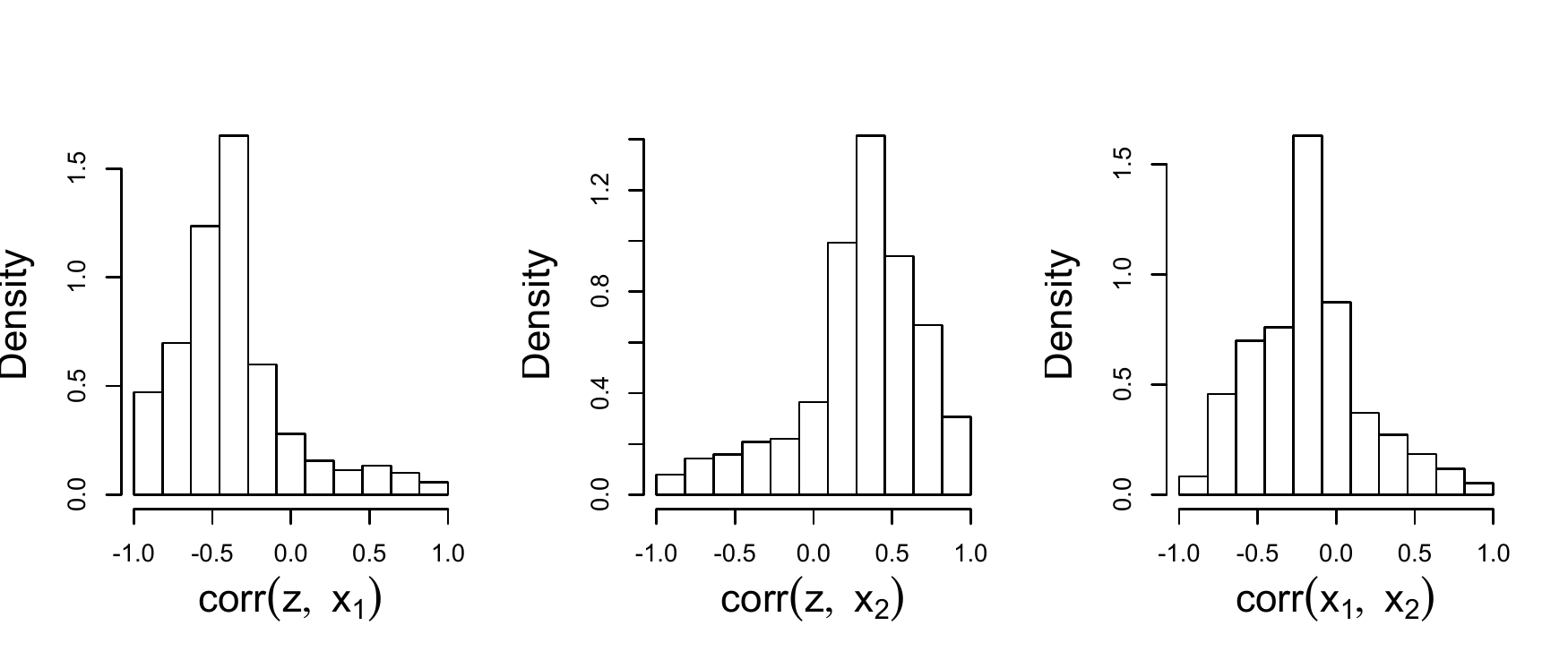}
\end{tabular}
\caption{Song sparrows data. Posterior predictive samples for 
$\mathrm{corr}(z,x_1)$ (left panel), $\mathrm{corr}(z,x_2)$ (middle
panel), and $\mathrm{corr}(x_1,x_2)$ (right panel).}
\label{post_corr}
\end{figure}

Finally, as a means to check if a kernel with independent components
for $x$ and $y$ would be adequate for this data example, we study
in posterior predictive space the correlations between the latent response 
and the two traits. Denoting by $\Theta$ the vector comprising all
model parameters, the joint posterior predictive distribution is given
by $p(z,x\mid\mathrm{data})=$
$\int \sum_{l=1}^{N} p_{l} \mathrm{N}_{3}(z,x;\mu_{l},\Sigma_{l})
p(\Theta \mid \mathrm{data}) \, \mathrm{d}\Theta$, which requires sampling 
one of $(\Sigma_{1},\dots,\Sigma_{N})$ with probabilities $p_{1},\dots,p_{N}$ 
for each set of posterior samples. 
The correlations resulting from these posterior predictive draws for
the kernel covariance matrix are plotted in Figure \ref{post_corr}.
These results suggest that it would be restrictive to force uncorrelated 
mixture kernel components, since
the distribution of correlations for $(z,x_1)$ is right-skewed and centered 
on negative values, while that for $(z,x_2)$ is mainly focused on
positive values and left-skewed, a pattern which is consistent with
the shape of the estimated binary regression curves.

\section{Discussion} 

We have presented a flexible method for estimating the regression relationship between 
binary responses and continuous covariates, which is built from a DP mixture
model for the latent response-covariate distribution.  Identifiability was established for the 
parameters of the mixture kernel.  In order to impose the restriction which is necessary 
for identifiability, the covariance matrix of the normal kernel was reparamaterized in such 
a way that allows for viewing only part of the matrix as random, while retaining the desirable 
features of conjugacy. Full conditional distributions were derived for the random elements of 
the covariance matrix, providing the key component of an efficient Markov chain Monte Carlo 
algorithm for posterior simulation.
Two strategies for prior specification were discussed. The methodology was illustrated with 
two data examples that were chosen to indicate the practical utility of the modeling approach 
for problems in the environmental sciences and in population biology.
%

We discussed the special case of the model arising from ${\Sigma}^{zx}={0}$ in the 
mixture kernel, which has been previously proposed with the further restriction that $\Sigma^{xx}$ 
is diagonal \citep{dunson}. There, the simplicity of independence among covariates 
within mixture components was viewed as appealing, and the response was modeled as independent of the 
covariates within the kernel, resulting in what was termed a product-kernel.    
In a related approach, \citet{shahbaba} also build a model for the joint distribution $f(y,x)$, 
but do so by separately estimating $f(x)$ and $f(y\mid x)$, where the latter is assumed to be a multinomial 
logit model within a mixture component.  Due to the difficulties arising from estimation of full covariance 
matrices unless the inflexible inverse-Wishart is used as a prior, they too assume $x_{1},...,x_{p}$ to be 
independent within each component. This idea was generalized by \citet{hannah} to allow
any standard generalized linear model to take the place of the multinomial logit model.

The independence assumptions discussed above are, in general,
restrictive. The proposed justification is that because independence
is imposed only within each component, dependence arises when more
than one component is contained in the mixture. Therefore, 
the ability of product-kernel models to approximate the regression
relationship and the covariate distribution is enhanced through the
mixture.  However, in order to correctly capture the covariate
distribution and the dependence of $y$ on $x$ in complex problems,
there is need for models which allow for dependence within clusters.\citet{dunson} note that if interest centers on
quantifying dependence, then there is no need to introduce a response,
and the method for joint modeling can still be used in this case.  If
estimation of dependence is in fact the goal, this is clearly more
adequately achieved when random variables are allowed to depend on one
another through more than just clustering.
In this work, the introduction of latent variables and reparameterization of the
covariance matrix allow these assumptions to be relaxed.
%

The proposed modeling approach relies on the choice of the
multivariate normal distribution for the mixture kernel. This choice can 
accommodate essentially any type of continuous covariate, possibly 
through use of appropriate transformation.
It can also handle ordinal categorical covariates $x$ by incorporating in the
model associated continuous variables, $x_{c}$, such that $x$ arises 
from $x_{c}$ through discretization. In particular, although in this
case inferences were not affected, beak length in the data example of 
Section 3.2 was recorded only to the nearest tenth, and it could 
therefore be treated as a discrete covariate.

%
%
%
%

This work lays in place the foundations for a variety of extensions to
ordinal regression problems involving data of
different types. In particular, extensions of the modeling approach to 
incorporate ordinal and mixed ordinal-continuous responses follow naturally.  
In analogy with the binary setting, a univariate ordinal response $y$ may be 
thought to arise as a discretized version of an underlying continuous response
$z$, such that $y=k$ if and only if 
$\gamma_{k-1} < z \leq \gamma_{k}$, for $k=2,...,K-1$, and $y=1$ or $y=K$ 
if and only if $z \leq \gamma_{1}$ or $z > \gamma_{K-1}$.
A normal DP mixture model can again be used for $(z,x)$.
However, extending the argument in \citet{kottas}, it can be
shown that if $K \geq 3$ all elements of the kernel covariance matrix are
identifiable when the cut-off points, $\gamma_{1},...,\gamma_{K-1}$,
are fixed. A key feature of the nonparametric mixture modeling framework 
is that we can obtain general inference with fixed cut-off points,
resulting in a great advantage over parametric models, the implementation
of which involves computationally challenging cut-off point estimation.
In the case of multivariate ordinal regression, 
each response may be assumed to arise from its own
underlying continuous response. Modeling these latent continuous
responses jointly with the covariates in the kernel sets the stage for
flexible inference on the relationship between the multivariate
ordinal response and the covariates, as well as among the ordinal responses.
Finally, we can consider mixed ordinal-continuous responses,
using a multivariate normal kernel for the latent responses, continuous 
responses, and covariates.  
%
We will report on these modeling extensions in a future manuscript.

\vspace{0.5cm}

\section*{Acknowledgements}
This research was supported in part by the National Science Foundation under award DMS 1310438.

\vspace{.5cm}

\appendix

\section*{Appendix A: Proofs of Lemmas 1 and 2}
\label{app::app_proofs}

\subsection*{Proof of Lemma 1}
Recall the kernel distribution in (\ref{eqn:induced}) for which we wish to prove that parameters
$(\mu^{x},\mu^{z},\Sigma^{xx},\Sigma^{zx})$ are identifiable, fixing $\Sigma^{zz}=1$. 
Assume that 
\begin{equation}
\label{eqn:equality}
k(y,x;\mu_{1}^{x},\mu_{1}^{z},\Sigma_{1}^{xx},\Sigma_{1}^{zx}) = 
k(y,x;\mu_{2}^{x},\mu_{2}^{z},\Sigma_{2}^{xx},\Sigma_{2}^{zx}).
\end{equation} 
If this implies $(\mu_{1}^{x},\mu_{1}^{z},\Sigma_{1}^{xx},\Sigma_{1}^{zx})=$
$(\mu_{2}^{x},\mu_{2}^{z},\Sigma_{2}^{xx},\Sigma_{2}^{zx})$, then 
$(\mu^{x},\mu^{z},\Sigma^{xx},\Sigma^{zx})$ are identifiable.

From (\ref{eqn:equality}), it must be the case that $\mathrm{N}_{p}(x;\mu_{1}^{x},\Sigma_{1}^{xx})=\mathrm{N}_{p}(x;\mu_{2}^{x},\Sigma_{2}^{xx})$.  This follows from summing each side of (\ref{eqn:equality}) over the two possible values of $y$.  Because the mean vector and covariance matrix are identifiable for the multivariate normal likelihood, it can be concluded that $\mu_{1}^{x}=\mu_{2}^{x}$, and $\Sigma_{1}^{xx}=\Sigma_{2}^{xx}$.  Now, after this simplification, each side of the equality in (\ref{eqn:equality}) consists of a Bernoulli distribution for $y\mid x$, and since $y$ is either 0 or 1, the corresponding Bernoulli probabilities must be equal.  Since $\Phi$ is a monotonically increasing function of its argument, the arguments of $\Phi$  are equal, that is,
\[\frac{\mu_{1}^{z}+\Sigma_{1}^{zx}(\Sigma^{xx})^{-1}(x-\mu^{x})}{(1-\Sigma_{1}^{zx}(\Sigma^{xx})^{-1}(\Sigma_{1}^{zx})^{t})^{1/2}}=\frac{\mu_{2}^{z}+\Sigma_{2}^{zx}(\Sigma^{xx})^{-1}(x-\mu^{x})}{(1-\Sigma_{2}^{zx}(\Sigma^{xx})^{-1}(\Sigma_{2}^{zx})^{t})^{1/2}}.\]
This can be written in the form ${a}^{t}x+b=0$, and in order for this to be true for all $x$, each element of vector ${a}$ must be 0, and scalar $b$ must be 0.  The two equations ${a}={0}$ and $b=0$ require

\begin{equation}\label{eqn:equality1}\frac{\Sigma_{1}^{zx}}{(1-\Sigma_{1}^{zx}(\Sigma^{xx})^{-1}(\Sigma_{1}^{zx})^{t})^{1/2}}=\frac{\Sigma_{2}^{zx}}{(1-\Sigma_{2}^{zx}(\Sigma^{xx})^{-1}(\Sigma_{2}^{zx})^{t})^{1/2}}\end{equation}
\begin{equation}\label{eqn:equality2}\frac{\mu_{1}^{z}-\Sigma_{1}^{zx}(\Sigma^{xx})^{-1}\mu^{x}}{(1-\Sigma_{1}^{zx}(\Sigma^{xx})^{-1}(\Sigma_{1}^{zx})^{t})^{1/2}}=\frac{\mu_{2}^{z}-\Sigma_{2}^{zx}(\Sigma^{xx})^{-1}\mu^{x}}{(1-\Sigma_{2}^{zx}(\Sigma^{xx})^{-1}(\Sigma_{2}^{zx})^{t})^{1/2}}\end{equation}
Using (\ref{eqn:equality1}), (\ref{eqn:equality2}) can be replaced by $\mu_{1}^{z}\Sigma_{2}^{zx}=\mu_{2}^{z}\Sigma_{1}^{zx}$. Writing these two equations component-wise, and letting $\Sigma_{ji}^{zx}$ denote element $i$ of the vector $\Sigma_{j}^{zx}$, results in two systems of $p$ equations:
\begin{equation}\label{eqn:system1}\frac{(\Sigma_{1i}^{zx})^{2}}{1-\Sigma_{1}^{zx}(\Sigma^{xx})^{-1}(\Sigma_{1}^{zx})^{t}}=\frac{(\Sigma_{2i}^{zx})^{2}}{1-\Sigma_{2}^{zx}(\Sigma^{xx})^{-1}(\Sigma_{2}^{zx})^{t}},\quad i=1,...,p\end{equation}
\begin{equation}\label{eqn:system2}\mu_{1}^{z}\Sigma_{2i}^{zx}=\mu_{2}^{z}\Sigma_{1i}^{zx},\quad i=1,...,p\end{equation}

When $p=1$ such that $\Sigma^{zx}$ is a scalar, (\ref{eqn:system1}) 
becomes $|\Sigma_{1}^{zx}|=|\Sigma_{2}^{zx}|$, which has only the
solution $\Sigma_{1}^{zx}=\Sigma_{2}^{zx}$, since $\Sigma_{1}^{zx}=-\Sigma_{2}^{zx}$ 
would violate (\ref{eqn:equality1}). Then from (\ref{eqn:system2}) we conclude $\mu_{1}^{z}=\mu_{2}^{z}$.  

In general, with $p$ covariates, (\ref{eqn:system1}) can be written as
\[(\Sigma_{1i}^{zx})^{2}-(\Sigma_{1i}^{zx})^{2}\sum_{k=1}^{p}\sum_{j=1}^{p}\Sigma_{2j}^{zx}\Sigma_{2k}^{zx}(\Sigma^{xx})_{jk}^{-1}=(\Sigma_{2i}^{zx})^{2}-(\Sigma_{2i}^{zx})^{2}\sum_{k=1}^{p}\sum_{j=1}^{p}\Sigma_{1j}^{zx}\Sigma_{1k}^{zx}(\Sigma^{xx})_{jk}^{-1},\quad i=1,...,p\]
Because (\ref{eqn:system2}) implies $\Sigma_{1l}^{zx}\Sigma_{2m}^{zx}=\Sigma_{1m}^{zx}\Sigma_{2l}^{zx}$ for any $l,m=1,...,p$, the equation reduces to $(\Sigma_{1i}^{zx})^{2}=(\Sigma_{2i}^{zx})^{2}$.  The constraint $\Sigma_{1l}^{zx}\Sigma_{2m}^{zx}=\Sigma_{1m}^{zx}\Sigma_{2l}^{zx}$ leaves only $\Sigma_{1}^{zx}=-\Sigma_{2}^{zx}$ and $\Sigma_{1}^{zx}=\Sigma_{2}^{zx}$ as possible solutions.  The first can be eliminated as well, since this contradicts (\ref{eqn:equality1}).  This leaves as the only feasible solution $\Sigma_{1}^{zx}=\Sigma_{2}^{zx}$, which implies $\mu_{1}^{z}=\mu_{2}^{z}$ from (\ref{eqn:system2}).  

It has been shown that if $k(y,x;\mu_{1}^{x},\mu_{1}^{z},\Sigma_{1}^{xx},\Sigma_{1}^{zx})=k(y,x;
\mu_{2}^{x},\mu_{2}^{z},\Sigma_{2}^{xx},\Sigma_{2}^{zx})$, then this implies $(\mu_{1}^{x},\mu_{1}^{z},\Sigma_{1}^{xx},\Sigma_{1}^{zx})=(\mu_{2}^{x},\mu_{2}^{z},\Sigma_{2}^{xx},\Sigma_{2}^{zx})$.  Therefore, applying directly the definition, the parameters $(\mu^{x},\mu^{z},\Sigma^{xx},\Sigma^{zx})$ are identifiable in the kernel of the mixture.

\subsection*{Proof of Lemma 2}
Consider $y=(y_{1},...,y_{r}) | \mu,\beta,\Delta \sim \mathrm{N}_{r}(\mu,\beta^{-1}\Delta(\beta^{-1})^{t})$, 
such that the likelihood for $\beta$ is proportional to 
$\exp\{-(y-\mu)^{t}\beta^{t}\Delta^{-1}\beta(y-\mu)\}$. First, focus on determining the likelihood 
for $\tilde{\beta}$, a vector of length $q=r(r-1)/2$. Write $\beta(y-\mu)$ as $M(1,\tilde{\beta}^{t})^{t}$, 
for a matrix $M$, of dimension $r \times (q+1)$ which has row $i$
containing $i$ nonzero elements, the first being $(y_{i}-\mu_{i})$,
occurring in column 1, and the rest being $(y_{1}-\mu_{1}),...,(y_{i-1}-\mu_{i-1})$, 
occurring in columns $2+(i-1)(i-2)/2$ to $i+(i-1)(i-2)/2$. Then, the likelihood for $\tilde{\beta}$ can be written proportional to 
$\exp\{-(1,\tilde{\beta}^{t})M^{t}\Delta^{-1}M(1,\tilde{\beta}^{t})^{t}\}$. Let $C=M^{t}\Delta^{-1}M$.  
If there exists a symmetric, positive definite matrix $T$ and vector $d$ for which 
$(1,\tilde{\beta}^{t})C(1,\tilde{\beta}^{t})^{t}=$ $\tilde{\beta}^{t}T\tilde{\beta}-2\tilde{\beta}^{t}Td+R$, 
where $R$ is a constant that does not depend on $\tilde{\beta}$, then the likelihood for $\tilde{\beta}$ 
corresponds to a normal distribution with mean vector $d$ and covariance matrix $T^{-1}$.  
The left side of the above equation is
$C_{11}+2\sum_{j=2}^{q+1}\tilde{\beta}_{j-1}C_{1j}+\sum_{j=2}^{q+1}\sum_{i=2}^{q+1}\tilde{\beta}_{j-1}\tilde{\beta}_{i-1}C_{ij}$,
and the last of these terms is just 
$\tilde{\beta}^{t} C_{q \times q} \tilde{\beta}$, where $C_{q\times q}$ denotes the $q\times q$
submatrix of $C$ obtained by deleting the first row and column of $C$.
Therefore, with $T=C_{q\times q}$, we seek $d$ such that 
$-\tilde{\beta}^{t}Td=\sum_{j=2}^{q+1}\tilde{\beta}_{j-1}C_{1j}$. Equating
the coefficient associated with $\tilde{\beta}_{i}$, $i=1,...,q$, on
each side of the equation results in a system of $q$ equations:
\begin{equation}
\label{eqn:meaneqn}
-\sum_{j=1}^{q}d_{j}T_{i-1,j}=C_{1i},\quad i=2,...,q+1.
\end{equation}
As explained in Section 2.2, $T$ is a block diagonal matrix which 
can be constructed from square matrices $T^{1},...,T^{r-1}$, of dimensions $1,...,r-1$, where 
\begin{equation}
\label{eqn:Telements}
T^{j}_{mn}=(y_{m}-\mu_{m})(y_{n}-\mu_{n})/\delta_{j+1},\quad m=1,...,j,\quad n=1,...,j.
\end{equation}  
The symmetry of $T$ follows from the symmetry of $C$, but it remains to be shown that $T$ is positive definite. For a non-zero vector $v$, we must have $v^{t}Tv>0$. When $r=2$, $v^{t}Tv$ becomes $v_{1}^{2}(y_{1}-\mu_{1})^{2}/\delta_{2}$. When $r=3$, $v^{t}Tv$ is the sum of the result for $r=2$ and the term  $(v_{2}(y_{1}-\mu_{1})+v_{3}(y_{2}-\mu_{2}))^{2}/\delta_{3}$. For $r=4$, the term $(v_{4}(y_{1}-\mu_{1})+v_{5}(y_{2}-\mu_{2})+v_{6}(y_{3}-\mu_{3}))^{2}/\delta_{4}$ is added to the result for $r=3$. In general, a term of the form $(v_{q-r+2}(y_{1}-\mu_{1})+...+v_{q}(y_{r-1}-\mu_{r-1}))^{2}/\delta_{r}$ is added in going from $r-1$ to $r$ dimensions. Clearly, $T$ is positive semidefinite. However, to have $v^{t}Tv>0$, and all elements of $T$ strictly positive, it must 
be the case that $y_{i}\neq\mu_{i}$, for $i=1,...,r-1$, which holds true with probability $1$, since $\mu$ is a continuous 
random vector.

We now derive the form of the mean vector $d$. Because $T$ is sparse, the system of $q$ equations (\ref{eqn:meaneqn}) can be divided into $r-1$ sets of equations, where set $j$ consists of $j$ equations with $j$ unknowns, $d_{1+j(j-1)/2},...,d_{j(j+1)/2}$.  Let the index $1+j(j-1)/2$ be denoted by $(1)$ and let the index $j(j+1)/2$ be denoted by $(j)$.  Set the first $j-1$ of these elements equal to 0, so that $d_{1+j(j-1)/2}=...=d_{j(j+1)/2-1}=0$.  Then the $j$ equations become \begin{equation}\label{eqn:jeqns}-d_{(j)}T_{(1),(j)}=C_{1,(1)+1},...,-d_{(j)}T_{(j),(j)}=C_{1,(j)+1}.\end{equation}  
The solution $d_{(j)}=-(y_{j+1}-\mu_{j+1})/(y_{j}-\mu_{j})$ satisfies these $j$ equalities (\ref{eqn:jeqns}), since the elements $C_{1,(1)+1},...,C_{1,(j)+1}$ are $(y_{1}-\mu_{1})(y_{j+1}-\mu_{j+1})/\delta_{j+1},...,(y_{j}-\mu_{j})(y_{j+1}-\mu_{j+1})/\delta_{j+1}$, and the elements $T_{(1),(j)},...,T_{(j),(j)}$ are $(y_{1}-\mu_{1})(y_{j}-\mu_{j})/\delta_{j+1},...,(y_{j}-\mu_{j})(y_{j}-\mu_{j})/\delta_{j+1}$, as given in (\ref{eqn:Telements}), so that \[-C_{1,(1)+1}/T_{(1),(j)}=...=-C_{1,(j)+1}/T_{(j),(j)}=-(y_{j+1}-\mu_{j+1})/(y_{j}-\mu_{j}).\]

With $n$ data vectors, $(y_{i,1},...,y_{i,r})$, for $i=1,...,n$, the likelihood for $\tilde{\beta}$ is proportional 
to a normal with mean $(\sum_{i=1}^{n}T_{i})^{-1}(\sum_{i=1}^{n}T_{i}d_{i})$, and covariance matrix 
$(\sum_{i=1}^{n}T_{i})^{-1}$, where $T_{i}$ and $d_{i}$ are computed using the $i$-th observation.  
When combined with a normal prior for $\tilde{\beta}$, the full conditional is also normal.

Next, consider the likelihood for the $\delta_{k}$, which up to the proportionality constant is given by 
$\prod_{k=1}^{r}\delta_{k}^{-1/2} \exp\{-\mathrm{tr}(\beta^{t}\Delta^{-1}\beta(y-\mu)(y-\mu)^{t})/2\}$. 
By properties of trace, $\mathrm{tr}(\beta^{t}\Delta^{-1}\beta(y-\mu)(y-\mu)^{t})=$ 
$\mathrm{tr}(\beta(y-\mu)(y-\mu)^{t}\beta^{t}\Delta^{-1})$. 
Let $A=$ $\beta(y-\mu)(y-\mu)^{t}\beta^{t}$.  Since $\Delta$ is diagonal with $\delta$ on the diagonal, 
the likelihood for each $\delta_{k}$ is proportional to $\delta_{k}^{-1/2} \exp\{-A_{kk}/(2 \delta_k)\}$.  
The diagonal elements of $A$ are the squares of $\beta(y-\mu)$, which are 
$A_{kk}=$ $\{ (y_{k}-\mu_{k}) + \sum_{j<k}\beta_{kj}(y_{j}-\mu_{j}) \}^{2}$. 
Then, with $n$ data vectors, $(y_{i,1},...,y_{i,r})$, $i=1,...,n$, the likelihood for $\delta_{k}$, $k=1,...,r$, 
is proportional to an inverse-gamma with shape parameter $(n/2)-1$ and scale parameter 
$0.5 \sum_{i=1}^{n} \{ (y_{i,k}-\mu_{k}) + \sum_{j<k}\beta_{kj}(y_{i,j}-\mu_{j}) \}^{2}$.  
When combined with an inverse-gamma prior, this results in a posterior
full conditional distribution which is inverse-gamma.

\section*{Appendix B: Distributions Implied by the inverse-Wishart}

Assume $\Sigma\sim \mathrm{IW}_{r}(v,T)$, with $r=p+1$, and partition $\Sigma$ into blocks, 
$\Sigma_{11}$, $\Sigma_{12}$, $\Sigma_{21}$, and $\Sigma_{22}$, of dimensions $q\times q$, $q\times (r-q)$, 
$(r-q)\times q$, and $(r-q)\times (r-q)$, respectively. Moreover, consider the corresponding partition for 
matrix $T$. Then, applying propositions 8.7 and 8.8 of Eaton (2007), we obtain:

(a) $\Sigma_{11}\sim \mathrm{IW}_q(v-(r-q),T_{11})$.

(b) $\Sigma_{22\cdot 1}\sim \mathrm{IW}_{r-q}(v,T_{22\cdot1})$, where $\Sigma_{22\cdot1}=$ $\Sigma_{22}-\Sigma_{21}\Sigma_{11}^{-1}\Sigma_{12}$ and $T_{22\cdot1}=$ $T_{22}-T_{21}T_{11}^{-1}T_{12}$.

(c) $\Sigma_{11}^{-1}\Sigma_{12} | \Sigma_{22\cdot1}^{-1}\sim \mathrm{MN}_{q,r-q}(T_{11}^{-1}T_{12},T_{11}^{-1},\Sigma_{22\cdot1})$. 
Here, MN denotes the matrix normal distribution such that, conditionally on $\Sigma_{22\cdot1}$, $\mathrm{vec}(\Sigma_{11}^{-1}\Sigma_{12})\sim$ 
$\mathrm{N}_{q(r-q)}(\mathrm{vec}(T_{11}^{-1}T_{12}),T_{11}^{-1}\otimes\Sigma_{22\cdot1})$.

We now assume $T$ is diagonal, with elements $(T_1,\dots,T_{p+1})$, as this is the case relevant to 
our prior specification approach. Let $T^i=$ $\mathrm{diag}(T_1,\dots,T_i)$. Applying result (b) with 
$q=p$, we obtain $\delta_{p+1}\sim \mathrm{IG}(0.5 v,0.5 T_{p+1})$. This uses the fact that 
$\Sigma_{22\cdot 1}=$ $\delta_{p+1}$ as a consequence of the $(\beta,\Delta)$ parameterization, and 
the simplification of $T_{22\cdot1}$ to $T_{22}=T_{p+1}$ when $T$ is diagonal. 
Applying result (a) with $q=p$, we obtain the marginal distribution of the upper left $p$ dimensional block 
of the covariance matrix $\Sigma$, which is $\Sigma_{1:p,1:p}\sim \mathrm{IW}_{p}(v-1,T^p)$. Next, using 
result (b) for matrix $\Sigma_{1:p,1:p}$ with $q=p-1$, we have $\delta_{p}\sim \mathrm{IG}(0.5 (v-1),0.5 T_{p})$, 
since $(\Sigma_{1:p,1:p})_{22\cdot1}=$ $\delta_{p}$. Analogously, applying results (a) and (b) in succession, 
we obtain $\delta_i\sim \mathrm{IG}(0.5 (v+i-(p+1)),0.5 T_{i})$, for $i=2,\dots,p+1$. 

For each $i=2,\dots,p+1$, result (a) yields an $\mathrm{IW}_i(v+i-(p+1),T^{i})$ distribution for 
$\Sigma_{1:i,1:i}$, that is, for the upper left block of $\Sigma$ of dimension $i$. Then, applying result (c) 
to $\Sigma_{1:i,1:i}$ with $q=i-1$, we obtain $(-\beta_{i,1},\dots,-\beta_{i,i-1})^t|\delta_i\sim$
$\mathrm{N}_{i-1}((0,\dots,0)^t,\delta_i(T^{i-1})^{-1})$, for $i=2,\dots,p+1$. This uses the fact that 
$(T^{i})_{12}=(0,\dots,0)^{t}$, $\mathrm{vec}( (\Sigma_{1:i,1:i})_{11}^{-1}(\Sigma_{1:i,1:i})_{12} )=$
$(-\beta_{i,1},\dots,-\beta_{i,i-1})^t$, and $(\Sigma_{1:i,1:i})_{22\cdot1} =$ $\delta_{i}$.

\section*{Appendix C: Model Comparison Criterion}

The predictive loss measure used for model comparison in Section 3.1 requires 
for each model $m$ the posterior predictive mean, $\mathrm{E}^{(m)}(y_{new,i}|\mathrm{data})$, 
and posterior predictive variance, $\mathrm{var}^{(m)}(y_{new,i}|\mathrm{data})$, for replicated 
response $y_{new,i}$ with associated covariate vector $x_i$.

Denote generically by $\Theta$ the full parameter vector for either the product-kernel model
or for the more general binary regression model developed in Section 2. For the former model, 
$$
\mathrm{E}(y|x_{i},\mathrm{data}) =
\{p(x_{i}|\mathrm{data})\}^{-1} \int\sum_{l=1}^{N} p_{l} \mathrm{N}_{p}(x_{i};\mu_{l}^{x},\Sigma_{l}^{xx})
\Phi(\mu_{l}^{z}) \, p(\Theta | \mathrm{data}) \mathrm{d}\Theta
$$
with $p(x_{i}|\mathrm{data})=$
$\int\sum_{l=1}^{N} p_{l} \mathrm{N}_{p}(x_{i};\mu_{l}^{x},\Sigma_{l}^{xx}) \, p(\Theta | \mathrm{data}) \mathrm{d}\Theta$, and
$\mathrm{E}(y^2|x_i,\mathrm{data})$ also has the same form. 
Under the proposed model, $\mathrm{E}(y|x_{i},\mathrm{data})$ is given by 
\[
\{p(x_{i}|\mathrm{data})\}^{-1}\int\sum_{l=1}^{N}p_{l} \mathrm{N}_{p}(x_{i};\mu_{l}^{x},\Sigma_{l}^{xx})
\Phi\left( \frac{\mu_{l}^{z}+{\Sigma_{l}^{zx}}({\Sigma_{l}^{xx}})^{-1}(x_{i} - {\mu_{l}^{x}})}
{(\Sigma_{l}^{zz}-{\Sigma_{l}^{zx}}({\Sigma_{l}^{xx}})^{-1}({\Sigma_{l}^{zx}})^{t})^{1/2}} \right)
\, p(\Theta | \mathrm{data}) \mathrm{d}\Theta
\]
where $p(x_i|\mathrm{data})=$
$\int\sum_{l=1}^{N}p_{l} \mathrm{N}_{p}(x_{i};\mu_{l}^{x},\Sigma_{l}^{xx}) \, p(\Theta | \mathrm{data}) \mathrm{d}\Theta$,
and $\mathrm{E}(y|x_{i},\mathrm{data})=$ $\mathrm{E}(y^2|x_{i},\mathrm{data})$. 
Hence, under both models, straightforward Monte Carlo integration using the posterior samples 
for model parameters yields estimates for the required posterior predictive means and variances.

\vspace{0.5cm}

\bibliographystyle{asa}
\bibliography{Binary_BA_bib}

\end{document}